\documentclass[aps,twocolumn,showpacs,pra,groupedaddress,superscriptaddress,nofootinbib,longbibliography]{revtex4-1}
\usepackage{graphicx,amsmath,amssymb,amsbsy,subfigure,hyperref,bbm,times,txfonts}
\usepackage{color}
\usepackage{makecell}
\begin{document}
\title{Steady-state coherence in multipartite quantum systems: 
its connection with thermodynamic quantities and impact on quantum thermal machines}

\author{Rui Huang}
\affiliation{School of Physics and Physical Engineering, Shandong Provincial Key Laboratory of Laser Polarization and Information Technology, Qufu Normal University, 273165, Qufu, China}

\author{Qing-Yu Cai}
\affiliation{School of Physics and Physical Engineering, Shandong Provincial Key Laboratory of Laser Polarization and Information Technology, Qufu Normal University, 273165, Qufu, China}

\author{Farzam Nosrati}
\affiliation{Dipartimento di Ingegneria, Università degli Studi di Palermo, Viale delle Scienze, 90128 Palermo, Italy}
\affiliation{IMDEA Networks Institute, Madrid, Spain}

\author{Rosario Lo Franco}%
\affiliation{Dipartimento di Ingegneria, Università degli Studi di Palermo, Viale delle Scienze, 90128 Palermo, Italy}

\author{Zhong-Xiao Man}\thanks{Corresponding author. Email: zxman@qfnu.edu.cn}
\affiliation{School of Physics and Physical Engineering, Shandong Provincial Key Laboratory of Laser Polarization and Information Technology, Qufu Normal University, 273165, Qufu, China}

\begin{abstract}
Understanding how coherence of quantum systems affects thermodynamic quantities, such as work and heat, is essential for harnessing quantumness effectively in thermal quantum technologies.	
Here, we study the unique contributions of quantum coherence among different subsystems of a multipartite system, specifically in non-equilibrium steady states, to work and heat currents.	
Our system comprises two coupled ensembles, each consisting of $N$ particles, interacting with two baths of different temperatures, respectively.
The particles in an ensemble interact with their bath either simultaneously or sequentially, leading to non-local dissipation and enabling the 
decomposition of work and heat currents into local and non-local components.
We find that the non-local heat current, as well as both the local and non-local work currents,
are linked to the system quantum coherence. 
We provide explicit expressions of coherence-related quantities
that determine the work currents under various intrasystem interactions.
Our scheme is versatile, capable of functioning as a refrigerator, an engine, and an accelerator, with its performance being highly sensitive to the configuration settings. These findings establish a connection between thermodynamic quantities and quantum coherence, supplying valuable insights for the design of quantum thermal machines.\\

\end{abstract}

\maketitle

\section{Introduction}
The rapid development of quantum science and technology \cite{QI} has prompted researchers to further explore the combination of quantum mechanics and other traditional fields. In this context, quantum thermodynamics has attracted extensive studies, with the aim to exploit quantum advantage in thermodynamic processes \cite{QT1,QT2,QT3,QT4,QT5,QT6}.
A topic of great interest is to use quantum resources to design quantum thermal machines (QTMs) that exhibit a better performance over their classical counterparts.
Experimental implementations of QTMs have utilized several platforms, such as trapped ions \cite{exp1,exp2,exp3,exp4}, trapped superfluid gas of atoms \cite{Koch2023}, and nuclear spin systems \cite{exp5,exp6}.
To effectively utilize quantum resources in thermodynamics, it is important to gain a comprehensive understanding of how traits such as quantum coherence within a quantum system impact thermodynamic quantities.

Recent years have seen notable advancements
in the study of boundary-driven quantum systems, where
the systems are coupled, at their boundaries,
to different baths \cite{boundary}.
The temperature gradient between the baths induces a heat flow
from one bath to the other, which constitutes the basic model 
for realizing QTMs.
In the stationary regime, the systems attain
a state known as the non-equilibrium steady-state (NESS) which potentially contains diverse quantum resources, such as coherence, that
have been shown to play a critical role in influencing the performance of QTMs
\cite{enh1,enh2,coh17,coh18,fric1,fric2,fric3,lubr,enh6,enh7,enh8,enh9,enh10,enh11,enh12,enh13}.
It has been demonstrated that small self-contained quantum refrigerators
composed of three qubits can surpass classical refrigerators in cooling capacity and energy transport 
achieved by the entanglement of the qubits \cite{enh1}.
Quantum engines that utilize two interacting qubits to execute a generalized Otto cycle
have been shown to exhibit a connection between the produced work and the correlations between the qubits \cite{enh2}.
Coherence in the working substance can produce power outputs
that significantly exceed that of equivalent classical machines \cite{coh17}.
An experiment using an ensemble of nitrogen vacancy centers in diamond has demonstrated the benefit of quantum coherence in providing power for QTMs \cite{coh18},
particularly in the so-called small action limit \cite{coh17}.
In some cases, quantum coherence in the working substance is associated with the
occurrence of quantum friction \cite{fric1,fric2,fric3}, although by carefully adjusting the parameters of the machines,
it can also act as a dynamical quantum lubricant \cite{lubr}.

In the stationary regime, coherence in a quantum system can be generated
via different mechanisms.
In the case of independent (or local) dissipation, occurring when the subsystems  are individually coupled to independent baths, 
interactions between subsystems are necessary to generate coherence. Independent dissipation is a popular model in designing QTMs
\cite{ManPRE17,multiind2,multiind3,boundary,Reconc,Smallest}.
Differently, non-local dissipation, emerging when
different transition levels or multiple subsystems are connected collectively to the same bath, can induce coherence even in the absence of
direct interactions between subsystems. 
It not only affects the energy exchange
between the system and the bath,
but also leads to richer forms of quantum coherence within the system.
The effects of non-local dissipation have been widely applied in quantum information technologies
and are now being used to improve the performance of quantum thermal machines
\cite{enh2,enh11,enh12,enh13,common1,common2,common3,common4,common6,common7,common8,common9,common10,common11}.
Moreover, non-local dissipation can arise in the so-called cascaded model \cite{cascaded1,cascaded2,cascaded3,cascaded4}, where
the subsystems, denoted as $S_1, S_2,\ldots,S_{N} $, of a multipartite system
interact with a bath one after the other in a cascaded manner, starting from $S_1$ until $S_N$.
Although both models can lead to non-local dissipation,
the system exhibits different dynamics in these two cases.
Notably, the cascaded model can cause unidirectional effects in the dynamics
between subsystems. As far as we know, the role of this one-way
effect on thermodynamic quantities has remained unexplored. In addition,  similarities and differences in the dependence of thermodynamic quantities on quantum coherence, under simultaneous or cascaded system-bath interactions, are worth to be investigated to characterize the two models and their impact on QTMs.

QTMs are usually modeled as open quantum systems in contact
with surrounding thermal baths, therefore the starting point
for the study is the theory of open quantum system.
The primary tool for describing the dynamics of the system
is quantum master equation (QME).
However, it has been shown that using QME to deal with quantum thermodynamics may result in thermodynamic inconsistency \cite{fric3,inconsis1,inconsis2,inconsis21,inconsis4,inconsis5,inconsis6}.
In addition to QME, the collision model (CM) is an efficient microscopic framework
to simulate the dynamics of an open quantum system \cite{CM,CM1,CM2,CM3,CM4,CM5,CM6,CM7,CM8,CM9}.
The CM describes the environment as a collection of identically prepared ancillas which individually interact, or collide, with the system at each time step.
The ancilla after the collision is discarded and a new one is introduced in the
next step. Recently, the CM has been employed in the exploration of quantum thermodynamics \cite{CMTher1,
CMTher4,CMTher5,CMTher6,CMTher7,CMTher8,CMTher10,CMTher11,CMTher12,CMTher13,CMTher9,
CMTher14,CMTher15,CMTher17}.
The CM can take environmental degrees of freedom into account,
making it convenient to identify all the involved energy costs in the thermodynamic process. It also permits to track the flow of information and energy between
the system and the environment.

In this study, by means of the framework of CM,
we unveil how the steady-state coherence within a
quantum system influences work and heat currents, establishing a
connection between them, which may be useful to enhance the performance of QTMs.
We consider both simultaneous and
cascaded system-bath interaction, analyzing different
forms of intrasystem interactions.
Due to the existence of non-local dissipation, both work and heat currents
can be decomposed into local and non-local components.
We observe that local and non-local work currents, as well as non-local heat current, are all influenced by the quantum coherence of the system.
However, these thermodynamic quantities depend on the coherence
of specific subsystems: non-local heat current is related to particles within the same bath, local work current relies on the coherence of particles in different baths,
while non-local work current is associated with both scenarios.
We also identify the coherence-related quantities that determine the
work currents in various configurations.
Our scheme can function as refrigerator, engine, or accelerator,
with its performance being highly sensitive to the specific configuration settings employed.

The paper is organized as follows. In Sec.~\ref{II} we describe the general scheme for both simultaneous and cascaded system-bath interaction models, and derive the QMEs for the system's dynamics.
In Sec.~\ref{III}, using the CM, we obtain heat and work currents
for both models, highlighting the relationship between the work currents and the associated measures of quantum coherence.
Sec.~\ref{IV} addresses the operating regimes and performance of our scheme as QTMs under different scenarios.
In Sec.~\ref{V} we provide our conclusions.

\section{Models and master equations}\label{II}

Our model considers a system made of two many-particle ensembles,
each coupled to a thermal bath.
For the interaction model between the system and the bath,
we consider two scenarios: in the first, all particles within the ensemble
interact with the bath simultaneously; in the second,
they interact with the bath in a sequential manner (see Fig.~\ref{MD}).
These are referred to as the simultaneous system-bath interaction model
and the cascaded system-bath interaction model, respectively.
Both of these situations can induce coherence between particles in the same ensemble,
even without direct interactions between them.
Furthermore, we also take into account various direct interactions between particles,
as well as the coherence consequently generated,
across the two ensembles. This general scenario allows a comprehensive characterization of the system. 

We use a local QME that is derived through
the CM method to describe the dynamics of our system.
Within the framework of CM, the bath is simulated as a collection of ancillas,
each prepared in the same thermal state.
The system interacts (collides) with one such bath ancilla for a short duration $\tau$,
after which the ancilla is replaced by a new one and this process is repeated sequentially.
In the continuous-time limit with $\tau\rightarrow0$,
a local QME in the Lindblad form can be derived.

In the following, we describe the models and construct the corresponding master equations for the system's dynamics.


\subsection{Simultaneous system-bath interaction model}\label{IIA}

\begin{figure}[h]
\begin{center}
{\includegraphics[width=0.99\linewidth]{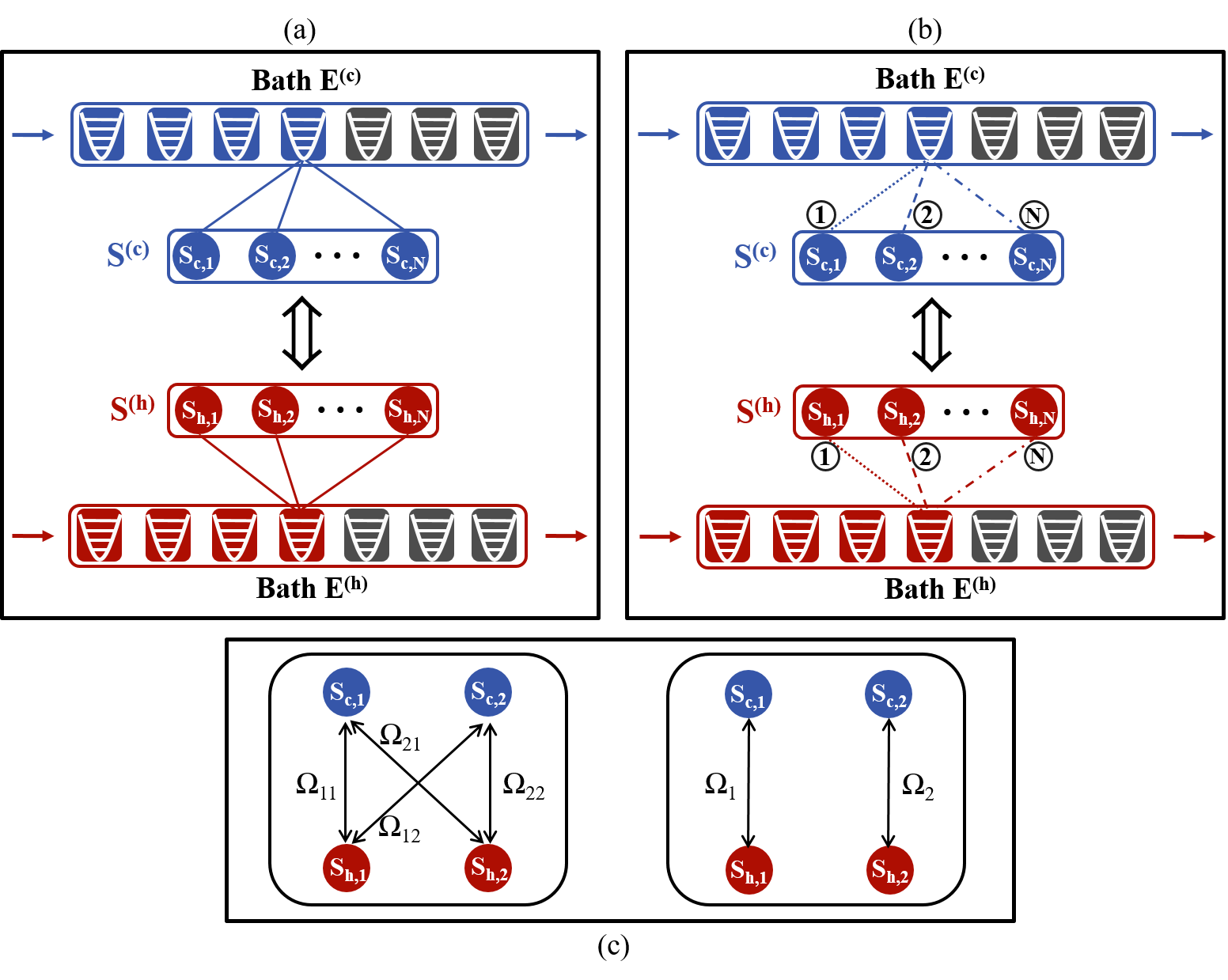}}
\end{center}
\vskip-0.5cm
\caption{Schematic representations of the common bath model (a) and the cascaded model (b).
In both models, the system comprises two ensembles $S^{(c)}$ and
$S^{(h)}$ of particles coupled to the baths $E^{(c)}$ and $E^{(h)}$, respectively.
By the framework of CM, the baths are simulated by a series of ancillas,
such as the harmonic oscillators shown in the figure.
In the common bath model (a), the $N$ particles in an ensemble interact
simultaneously with the bath ancilla, while in the cascaded model (b),
the particles of an ensemble interact with the bath ancilla in a definite order, labeled by
the notations \textcircled{\scriptsize{1}}, \textcircled{\scriptsize{2}},
\ldots,\textcircled{\scriptsize{N}}.
We consider two types of inner interactions between the
two ensembles, which is illustrated in panel (c) for the case of $N=2$.
In the first type of intrasystem interaction (left figure of panel (c)),
the particle in one ensemble interacts with all the particles in the other ensemble,
whereas in the second type of interaction (right figure panel (c)),
the particle in one ensemble only interacts with the corresponding particle in sequence
in the other ensemble. }
\label{MD}
\end{figure}

In this section, we address the simultaneous system-bath interaction
model, depicted in Fig.~\ref{MD}(a), where the particles in an ensemble
are simultaneously coupled to the bath and indistinguishable from the perspective of the bath.
The Hamiltonian of the system is given as
\begin{equation}
\hat{H}_{S}=\sum_{i=h,c}\sum_{n=1}^{N}\hat{H}_{S_{i,n}}+\hat{H}_{I},
\end{equation}
where $\hat{H}_{S_{i,n}}$ is the Hamiltonian of the $n$-th subsystem $S_{i,n}$
in the ensemble $S^{(i)}$ and $\hat{H}_{I}$
accounts for interactions of subsystems across the two ensembles.
The total Hamiltonian regarding the system, the bath ancilla and the system-bath interaction
can be summarized as
\begin{equation}\label{Htot}
\hat{H}_{tot}=\hat{H}_{S}+\sum_{i=h,c}\hat{H}_{E_i}+\frac{1}{\sqrt{\tau}}
\sum_{i=h,c}\sum_{n=1}^{N}\hat{V}_{i,n},
\end{equation}
where $\hat{H}_{E_i}$ represents the Hamiltonian of generic ancilla $E_{i}$ of the bath $E^{(i)}$
and $\hat{V}_{i,n}$ governs the collision between $S_{i,n}$ and
$E_i$. For the convenience of taking the continuous time limit, we have
scaled $\hat{V}_{i,n}$ by the collision duration $\tau$.
After undergoing a collision, the reduced state $\rho_{S}\equiv \rho_{S}(t)$ at time $t$ of $2N-$particle system,
made of $N$ subsystems in $S^{(c)}$ and $N$ subsystems in $S^{(h)}$,
is transformed into
$\rho_{S}^{\prime}\equiv \rho_{S}(t+\tau)$ determined by the map
\begin{equation}\label{mapcomm}
\begin{aligned}
\rho^{\prime}_{S}=\mathrm{Tr}_{E}\rho_{SE}^{\prime}
=\mathrm{Tr}_{E}\left[\hat{U}(\tau)\rho_{SE}\hat{U}^{\dag}(\tau)\right],
\end{aligned}
\end{equation}
with $\rho_{SE}=\rho_{S}\otimes\rho_{E_{h}}\otimes\rho_{E_{c}}$,
$\rho_{E_{h}}$ ($\rho_{E_{c}}$) the state of generic ancilla $E_{h}$ ($E_{c}$) of bath
$E^{(h)}$ ($E^{(c)}$), and
$\hat{U}(\tau)=e^{-i\tau \hat{H}_{tot}}$ the time evolution operator.
By expanding $\hat{U}(\tau)$ to a power series in $\tau$ up to its first order
and taking the limit $\tau\rightarrow0$,
we obtain the QME for the system's dynamics as
\begin{eqnarray}\label{QME}
\dot{\rho}_{S}&=&\lim_{\tau\rightarrow 0}\left[\left(\rho^{\prime}_{S}-\rho_{S}\right)/\tau\right]\nonumber\\
&=&-i\left[\hat{H}_{S},\rho_{S}\right]+\sum_{i=h,c}\mathcal{D}^{(i)}_{loc}\left(\rho_{S}\right)
+\sum_{i=h,c}\mathcal{D}^{(i)}_{non-loc}\left(\rho_{S}\right),
\end{eqnarray}
where $\mathcal{D}^{(i)}_{loc}\left(\rho_{S}\right)$
and $\mathcal{D}^{(i)}_{non-loc}\left(\rho_{S}\right)$ represent the local
and non-local dissipations being of the forms
\begin{equation}\label{Diloc}
\mathcal{D}^{(i)}_{loc}(\rho_{S})=-\frac{1}{2}\sum_{n=1}^{N}\mathrm{Tr}_{E}
\left[\hat{\widetilde{V}}_{i,n},\left[\hat{\widetilde{V}}_{i,n},\rho_{SE}\right]\right],
\end{equation}
and
\begin{equation}\label{Dinonloc}
\mathcal{D}^{(i)}_{non-loc}(\rho_{S})=-\frac{1}{2}\sum_{n'\neq n=1}^{N}\mathrm{Tr}_{E}\left[\hat{\widetilde{V}}_{i,n'},
\left[\hat{\widetilde{V}}_{i,n},\rho_{SE}\right]\right],
\end{equation}
where $\hat{\widetilde{V}}_{i,m}\equiv
I_{i,1} \otimes \cdots \otimes \hat{V}_{i,m} \otimes \cdots \otimes I_{i,N} \otimes \mathbf{I}_{i'}^{\otimes N}$
with $I_{i,m}$ the identity operator on particle $m$ in the bath $E^{(i)}$
and $\mathbf{I}_{i'}^{\otimes N}$ the global identity operator on the
particles in the bath $E^{(i')}$ ($i'\neq i$), and $\mathrm{Tr}_{E}$ denotes the
trace over all particles in both $E^{(h)}$ and $E^{(c)}$.
The local dissipation of the $n-$th subsystem
is equivalent to the situation where it interacts with the bath alone
in the absence of other subsystems.
By contrast, the non-local dissipations
describe collective energy exchanges between the subsystems
and the bath, which cannot be separated into contributions of individual subsystems.

To be specific, we consider the subsystems in the ensemble $S^{(i)}$ ($i=h,c$) as
$N$ identically prepared two-level systems (TLSs) with the generic Hamiltonian ($\hbar \equiv 1$)
\begin{equation}\label{HTLS}
\hat{H}_{S_{i,n}}=\omega_{i}\hat{\sigma}_{i,n}^{+}\hat{\sigma}_{i,n}^{-},
\end{equation}
in which $\omega_{i}$ is the transition frequency and $\hat{\sigma}_{i,n}^{\pm}=(\hat{\sigma}_{i,n}^{x}\pm i\hat{\sigma}_{i,n}^{y})/2$ with $\left\{\hat{\sigma}^{x},\hat{\sigma}^{y},\hat{\sigma}^{z}\right\}$
the usual Pauli operators.
The ancilla in the bath $E^{(i)}$ is taken to be harmonic oscillator with the Hamiltonian 
\begin{equation}\label{HHO}
\hat{H}_{E_{i}}=\omega_{i}\hat{a}_{i}^{\dag}\hat{a}_{i},
\end{equation}
with $\hat{a}_{i}^{\dag}$ ($\hat{a}_{i}$) the creation (annihilation) operator for the
oscillator.
The interaction between the TLS $S_{i,n}$ in
the ensemble $S^{(i)}$ with the bath ancilla $E_{i}$
is depicted as
\begin{equation}\label{Vi}
\hat{V}_{i,n}=g_{i,n}\left(\hat{\sigma}_{i,n}^{+}\hat{a}_{i}
+\hat{\sigma}_{i,n}^{-}\hat{a}_{i}^{\dag}\right),
\end{equation}
with $g_{i,n}$ the coupling strength.
By substituting the explicit form of $\hat{V}_{i,n}$ (\ref{Vi}) into Eqs.~(\ref{Diloc})
and (\ref{Dinonloc}), we can obtain
\begin{eqnarray}\label{Di2loc}
\mathcal{D}^{(i)}_{loc}(\rho_{S})
&=&\sum_{n=1}^{N}\gamma_{i,nn}^{-}\left[\hat{\widetilde{\sigma}}_{i,n}^{-}\rho_{S}
\hat{\widetilde{\sigma}}_{i,n}^{+}
-\frac{1}{2}\left\{\hat{\widetilde{\sigma}}_{i,n}^{+}\hat{\widetilde{\sigma}}_{i,n}^{-},
\rho_{S}\right\}\right]\nonumber\\
&+&\gamma_{i,nn}^{+}\left[\hat{\widetilde{\sigma}}_{i,n}^{+}\rho_{S}\hat{\widetilde{\sigma}}_{i,n}^{-}
-\frac{1}{2}\left\{\hat{\widetilde{\sigma}}_{i,n}^{-}\hat{\widetilde{\sigma}}_{i,n}^{+},\rho_{S}\right\}\right],
\end{eqnarray}
and
\begin{eqnarray}\label{Di2nonloc}
\mathcal{D}^{(i)}_{non-loc}(\rho_{S})
&=&\sum_{n^{\prime}\neq n=1}^{N}\gamma_{i,nn'}^{-}\left[\hat{\widetilde{\sigma}}_{i,n}^{-}\rho_{S}\hat{\widetilde{\sigma}}_{i,n'}^{+}
-\frac{1}{2}\left\{\hat{\widetilde{\sigma}}_{i,n}^{+}\hat{\widetilde{\sigma}}_{i,n'}^{-},
\rho_{S}\right\}\right]\nonumber\\
&+&\gamma_{i,nn'}^{+}\left[\hat{\widetilde{\sigma}}_{i,n}^{+}\rho_{S}\hat{\widetilde{\sigma}}_{i,n'}^{-}
-\frac{1}{2}\left\{\hat{\widetilde{\sigma}}_{i,n}^{-}
\hat{\widetilde{\sigma}}_{i,n'}^{+},\rho_{S}\right\}\right],
\end{eqnarray}
where
$\gamma_{i,kl}^{-}=g_{i,k}g_{i,l}\left\langle \hat{a}_{i}\hat{a}_{i}^{\dag}\right\rangle=g_{i,k}g_{i,l}\left(n_{i}+1\right)$,
$\gamma_{i,kl}^{+}=g_{i,k}g_{i,l}\left\langle \hat{a}_{i}^{\dag}\hat{a}_{i}\right\rangle=g_{i,k}g_{i,l}n_{i}$
with $n_{i}=1/(e^{\beta_{i}\omega_{i}}-1)$ the average photon number of the bath
at frequency $\omega_{i}$ and $\beta_{i}=1/T_{i}$ the bath's inverse temperature (we have set $k_{B}\equiv 1$), and
$\hat{\widetilde{\sigma}}_{i,m}^{\pm}\equiv I_{i,1} \otimes \cdots \otimes \hat{\sigma}_{i,m}^{\pm} \otimes \cdots \otimes I_{i,N} \otimes \mathbf{I}_{i'}^{\otimes N}$.
For the thermal bath $E^{(i)}$ at inverse temperature $\beta_{i}$,
the dissipation rates satisfy the local detailed balance
\begin{equation}
\frac{\gamma_{i,kl}^{+}}{\gamma_{i,kl}^{-}}=e^{-\beta_{i}\omega_{i}}.
\end{equation}

\subsection{Cascaded system-bath interaction model}\label{IIB}
Apart from simultaneous system-bath interactions,
the steady-state coherence can also be induced by
the cascaded system-bath interactions.
The cascaded model we consider still comprises
two particle ensembles, $S^{(h)}$ and $S^{(c)}$, that are in contact with two baths,
$E^{(h)}$ and $E^{(c)}$, respectively. Unlike the simultaneous interactions,
in the cascaded model the subsystems $S_{i,1},S_{i,2},\ldots,S_{i,N}$ ($i=h,c$) of the ensemble $S^{(i)}$
interact with the bath ancilla $E_{i}$ sequentially, as illistrated in Fig.~\ref{MD} (b).
This means that the interaction $S_{i,1}$-$E_{i}$ occurs first,
followed by $S_{i,2}$-$E_{i}$, $S_{i,3}$-$E_{i}$, and so on, 
until $S_{i,N}$-$E_{i}$.
Then, the above process is repeated such that the subsystems collide with the next ancilla.
Each collision lasts a duration $\tau$, thus the total time required for the
entire interactions from $S_{i,1}-E_{i}$ to $S_{i,N}-E_{i}$ is $N\tau$.
The state $\rho_{S}$ of the system at time $t$ is transformed into $\rho_{S}^{\prime}=\mathrm{Tr}_{E} [ \rho_{SE}^{\prime}]$ at time $t+N\tau$ by the map
\begin{equation}\label{mapcas}
\rho^{\prime}_{S}=\mathrm{Tr}_{E}\left[\hat{U}_{N}(\tau)\cdots \hat{U}_{2}\hat{U}_{1}(\tau)
\rho_{SE}\hat{U}_{1}^{\dag}(\tau)\hat{U}_{2}^{\dag}(\tau)\cdots \hat{U}^{\dag}_{N}(\tau)\right],
\end{equation}
where $\rho_{SE}=\rho_{S}\otimes\rho_{E_{h}}\otimes\rho_{E_{c}}$, with
$\rho_{E_{h}}$ and $\rho_{E_{c}}$ the states of bath ancillas $E_{h}$ and $E_{c}$, respectively.
The map Eq.~(\ref{mapcas}) for the cascaded model clearly reflects the order of interaction
between subsystems and the bath, which is significantly different from the map
(\ref{mapcomm}) for the simultaneous interaction model.
In the map (\ref{mapcas}), the unitary time-evolution operator
$\hat{U}_{n}(\tau)=e^{-i\tau\hat{H}_{n}}$,
where $\hat{H}_{n}=\sum_{i=h,c}\left(\hat{H}_{S_{i,n}}
+\hat{H}_{E_i}+\hat{V}_{i,n}+\hat{H}_{I,n}\right)$ with
$\hat{H}_{S_{i,n}}$ and $\hat{H}_{E_i}$ the free Hamiltonians of
$S_{i,n}$ and $E_{i}$, $\hat{V}_{i,n}$ the interacting Hamiltonian
between $S_{i,n}$ and $E_{i}$, and $\hat{H}_{I,n}$ capturing the interactions
between $S_{i,n}$ and the other subsystems.

After expanding $\hat{U}_{n}\left(\tau\right)$ as a power series and taking the continuous-time limit of
$\tau\rightarrow0$, we obtain the QME describing the system's evolution
in the same form as that given in Eq.~(\ref{QME}). The system's Hamiltonian
in the present case becomes
$\hat{H}_{S}=\sum_{n=1}^{N}\left(\hat{H}_{S_{h,n}}+\hat{H}_{S_{c,n}}+\hat{H}_{I,n}\right)$.
The local dissipation term $\mathcal{D}^{(i)}_{loc}(\rho_{S})$ is the same
as that given in Eq.~(\ref{Diloc}).
However, the non-local dissipation $\mathcal{D}^{(i)}_{non-loc}(\rho_{S})$
takes a completely different form compared to that in Eq.~(\ref{Dinonloc}) as
\begin{equation}\label{Dnoncas1}
\mathcal{D}^{(i)}_{non-loc}(\rho_{S})=-\sum_{n'>n=1}^{N}
\mathrm{Tr}_{E}\left[\hat{\widetilde{V}}_{i,n'},
\left[\hat{\widetilde{V}}_{i,n},\rho_{SE}\right]\right],
\end{equation}
which, although formally acting on both $S_{i,n}$ and $S_{i,n'}$, generally
results in an one-way impact of $S_{i,n}$ that collides with the bath beforehand
to $S_{i,n'}$ that collides with the bath afterwards.
This one-way effect is reflected mathematically in the subscript
of the summation notation, which is $n'>n$ here, whereas it is $n'\neq n$ in Eq.~(\ref{Dinonloc}).

We continue to examine the specific situation when the system and baths
are made up of TLSs and harmonic oscillators, as depicted in Eqs.~(\ref{HTLS})
and (\ref{HHO}), respectively, and the interaction Hamiltonian $\hat{V}_{i,n}$
between the $n$-th TLS in the ensemble $S^{(i)}$ and the bath ancilla $E_{i}$
takes the form of Eq.~(\ref{Vi}).
For this setting, the local dissipation term has the same form as that presented in
Eq.~(\ref{Di2loc}), while the non-local dissipation reads
\begin{eqnarray}\label{Dnoncas2}
\mathcal{D}^{(i)}_{non-loc}(\rho_{S})&=&\sum_{n'>n=1}^{N}\gamma_{i,nn'}^{-}
\left(\hat{\widetilde{\sigma}}_{i,n}^{-}
\left[\rho_{S},\hat{\widetilde{\sigma}}_{i,n'}^{+}\right]
+\left[\hat{\widetilde{\sigma}}_{i,n'}^{-},\rho_{S}\right]
\hat{\widetilde{\sigma}}_{i,n}^{+}\right)\nonumber\\
&+&\gamma_{i,nn'}^{+}\left(\hat{\widetilde{\sigma}}_{i,n}^{+}
\left[\rho_{S},\hat{\widetilde{\sigma}}_{i,n'}^{-}\right]
+\left[\hat{\widetilde{\sigma}}_{i,n'}^{+},\rho_{S}\right]\hat{\widetilde{\sigma}}_{i,n}^{-}\right),
\end{eqnarray}
where the expressions of $\gamma_{i,nn'}^{-}$ and $\gamma_{i,nn'}^{+}$
are identical to those given in Eq.~(\ref{Di2nonloc}).

\section{Work and heat currents}\label{III}
To develop QTMs and utilize quantum resources to enhance their
performance, it is essential to accurately determine thermodynamic
quantities and understand their properties. A previous study has derived
the heat and work for boundary-driven interacting systems
in independent baths (i.e., only local dissipation exists) \cite{Reconc}. In
this work, we aim to establish explicit formulations for heat and work
in more complex configurations and to uncover the effects of
steady-state coherence on them.
We consider both the simultaneous and
cascaded system-bath interactions that can induce non-local dissipation,
and compare the thermodynamic quantities under these two models.

\begin{table*}[htbp]
	\centering
	\caption{Table of relationship of thermodynamic quantities with steady-state coherence and population of the system. The check mark indicates that the thermodynamic quantity has a relationship
		with the corresponding factor. }
	\begin{tabular}{|c | c|c|c|}\hline
		\thead{\bf Thermodynamic quantities} & \thead{ \bf Population} & \thead{ \bf Coherence of subsystems \\ \bf  in the same bath}  & \thead{ \bf Coherence of subsystems\\ \bf  across two baths}  \\ \hline\hline
		\thead{local heat } & \thead{$\surd$} &       &   \\ \hline
		\thead{non-local heat}     &       & \thead{$\surd$} &  \\ \hline
		\thead{local work}         &       &       & \thead{$\surd$}  \\ \hline
		\thead{non-local work}     &       & \thead{$\surd$} & \thead{$\surd$}  \\ \hline
	\end{tabular}\label{table1}
\end{table*}

In terms of the first law of thermodynamics, the change rate of 
internal energy $U=\mathrm{Tr}(H_{S}\rho_{S})$ of the system
can be divided into work and heat currents, i.e., $\dot{W}$
and $\dot{Q}=\sum_{i=h,c}\dot{Q}_{i}$, in the sense of $\dot{U}=\dot{W}+\dot{Q}$.
Within the framework of CM, heat is defined unambiguously as the
energy change of bath ancilla.
The heat transferred from the bath $E_{i}$ to the system reads
\begin{equation}
\Delta Q_{i}=\left\langle \hat{H}_{E_{i}}\right\rangle_{\rho_{SE}}
-\left\langle \hat{H}_{E_{i}}\right\rangle_{\rho_{SE}^{\prime}},
\end{equation}
where $\left\langle\cdot\right\rangle_{\rho}\equiv\mathrm{Tr}\left(\cdot\rho\right)$.
The heat current regarding $E_{i}$ can correspondingly be defined as 
$\dot{Q}_{i}=\lim_{\tau\rightarrow0}\Delta Q_{i}/\tau$.
By contrast, the work is a very subtle quantity in quantum thermodynamics,
and its accurate identification and definition are
essential for retaining thermodynamic consistency and establishing QTMs.
Within the framework of CM, the successive coupling-decoupling between the system and bath
necessitates the input of work, and at the same time the internal interactions of the system also contribute to
the work. In this paper, we expect to control the work by manipulating intrasystem interactions so that
the work involved in this model should be solely due to the intrasystem interactions.
To this end, we assume that the system-bath coupling is energy conservation
in the sense of $\left[\sum_{n=1}^{N}\hat{H}_{S_{i,n}}+\hat{H}_{E_{i}},\sum_{n=1}^{N}\hat{V}_{i,n}\right]=0$ for $i=h,c$,
which means that the energy leaves the baths will enter the system and vice versa.
The inner interactions between particles of the two
ensembles, however, cannot maintain the global energy conservation with
$\left[\sum_{n=1}^{N}\hat{H}_{S_{i,n}}+\hat{H}_{I}+\hat{H}_{E_{i}},\sum_{n=1}^{N}\hat{V}_{i,n}
\right]=\left[\hat{H}_{I},\sum_{n=1}^{N}\hat{V}_{i,n}\right]\neq 0$.
Therefore, it is necessary to have external work source
coupled with the system in order to drive the system's dynamics.
During a collision process, the system and baths as a whole undergo unitary dynamical evolution,
hence the work involved in this process can be defined as
\begin{equation}
\Delta W=\int_{t}^{t+\tau}\left\langle\frac{\partial \hat{H}_{tot}}
{\partial s}\right\rangle_{\rho_{SE}} ds.
\end{equation}
In the collision model, the system undergoes successive coupling and decoupling from the
baths, the total Hamiltonian $\hat{H}_{tot}$ (\ref{Htot}) is actually time dependent.
That is, the interaction Hamiltonian $\hat{V}_{i,n}$ only exists in the interval 
of system-bath collisions and vanishes otherwise, which therefore 
is the only time-dependent term \cite{Reconc}. 
An integration over the above equation yields a general
formulation of work as,
\begin{equation}\label{DW}
\Delta W=\frac{1}{\sqrt{\tau}}\sum_{i=h,c}\sum_{n=1}^{N}\left[
\left\langle\hat{V}_{i,n}\right\rangle_{\rho_{SE}}
-\left\langle\hat{V}_{i,n}\right\rangle_{\rho_{SE}^{\prime}}\right].
\end{equation}
The work current can be given as
$\dot{W}=\lim_{\tau\rightarrow0}\Delta W/\tau$.

In the following, we shall first provide general formulations of work and heat
under both simultaneous and cascaded interactions between the system and the bath.
Due to the presence of non-local dissipation, both heat and work can be decomposed into
local and non-local components. By considering the particles in each ensemble as TLSs and
the bath ancillae as harmonic oscillators, and introducing different forms
of interaction between particles across the two ensembles,
we obtain specific expressions for work and heat in various scenarios.
Through these analytical expressions, we can determine the dependence
of thermodynamic quantities on the steady-state coherence or population of the system.
Moreover, we find coherence-related quantities that determine local and nonlocal work.

Before presenting detailed discussions, we summarize the relationships
of thermodynamic quantities with steady-state population and coherence of the system in
Table~\ref{table1}. It shows that local heat, non-local heat and local work
are determined by population, coherence of subsystems in the same bath and coherence of subsystems across two baths, respectively. In contrast, non-local work is related both to quantum coherence in the same bath and quantum coherence across two baths.

\subsection{Simultaneous system-bath interaction model}\label{IIIA}
\subsubsection{General formulations of work and heat currents}\label{IIIA1}
We first establish the relevant thermodynamic quantities, namely work and heat currents, for the
simultaneous system-bath interaction and decompose them into local and non-local parts (see Appendix \ref{app:WQ} for details). 

By taking the limit of $\tau\rightarrow0$, we obtain the work current
\begin{equation}\label{workcurr}
\dot{W}^{com}=\lim_{\tau\rightarrow 0}\left(\Delta W/\tau\right)=\dot{W}^{com}_{loc}+
\dot{W}^{com}_{non-loc},
\end{equation}
with
\begin{equation}\label{wcurrloc}
\dot{W}^{com}_{loc}=-\frac{1}{2}\sum_{i=h,c}\sum_{n=1}^{N}\left\langle\left[\hat{V}_{i,n},\left[\hat{V}_{i,n},
\hat{H}_{S}+\hat{H}_{E_{i}}\right]\right]\right\rangle_{\rho_{SE}},
\end{equation}
and
\begin{equation}\label{wcurrnloc}
\dot{W}^{com}_{non-loc}=-\frac{1}{2}\sum_{i=h,c}\sum_{n^{\prime}\neq n=1}^{N}
\left\langle\left[\hat{V}_{i,n'},\left[\hat{V}_{i,n},
\hat{H}_{S}+\hat{H}_{E_{i}}\right]\right]\right\rangle_{\rho_{SE}},
\end{equation}
its local and non-local components. Here, the superscript ``com"
in the work current indicates that the ensemble particles interact
simultaneously with a ``common" bath.
The local work current in Eq.~(\ref{wcurrloc})
is equivalent to the one derived in Ref.~\cite{Reconc}, which is applicable when
only independent dissipations are involved for the system.
Due to the presence of nonlocal dissipation in our model,
the system exhibits additional coherence in the steady state that
would not be present under independent dissipation, as will be seen in the
specific models discussed later.
As a result, an additional non-local work current $\dot{W}^{com}_{non-loc}$ arises,
as given in Eq.~(\ref{wcurrnloc}), which
provides an additional
channel for the system to exchange energy with external work source.

Similarly, we obtain the heat current associated with the bath $E^{(i)}$,
which can be decomposed into the local and non-local
parties as
\begin{equation}
\dot{Q}^{com(i)}=\lim_{\tau\rightarrow 0}\left(\Delta Q_{i}/\tau\right)
=\dot{Q}^{com(i)}_{loc}+\dot{Q}^{com(i)}_{non-loc},
\end{equation}
with
\begin{equation}\label{Qloc}
\dot{Q}^{com(i)}_{loc}=\frac{1}{2}\sum_{n=1}^{N}\left\langle
\left[\hat{V}_{i,n},\left[\hat{V}_{i,n},\hat{H}_{E_i}
\right]\right]\right\rangle_{\rho_{SE}},
\end{equation}
and
\begin{equation}\label{Qnonloc}
\dot{Q}^{com(i)}_{non-loc}=\frac{1}{2}\sum_{n^{\prime}\neq n=1}^{N}\left\langle\left[\hat{V}_{i,n'},\left[\hat{V}_{i,n},\hat{H}_{E_{i}}
\right]\right]\right\rangle_{\rho_{SE}}.
\end{equation}

\subsubsection{Specific model for the system and bath made up of TLSs and harmonic oscillators }
Returning to the situation where the system is composed of TLSs,
the bath ancilla is an harmonic oscillator, and system-bath interactions are governed by the operator $\hat{V}_{i,n}$ of Eq.~(\ref{Vi}),
the local and non-local heat currents become
\begin{equation}
\dot{Q}^{com(i)}_{loc}=\omega_{i}\sum_{n=1}^{N}\left\{\gamma_{i,nn}^{+}\left\langle\hat{\sigma}_{i,n}^{-}
\hat{\sigma}_{i,n}^{+}\right\rangle_{\rho_{S}}-\gamma_{i,nn}^{-}\left\langle\hat{\sigma}_{i,n}^{+}
\hat{\sigma}_{i,n}^{-}\right\rangle_{\rho_{S}}\right\},
\end{equation}
and
\begin{equation}\label{Qinonloc}
\dot{Q}^{com(i)}_{non-loc}=\frac{1}{2}\omega_{i}\sum_{n^{\prime}\neq n=1}^{N}\left(\gamma_{i,nn'}^{+}-\gamma_{i,nn'}^{-}\right)\left\langle\left(\hat{\sigma}_{i,n}^{+}
\hat{\sigma}_{i,n'}^{-}\right)_{+}\right\rangle_{\rho_{S}},
\end{equation}
with $\left(\hat{\sigma}_{\mu}^{+}\hat{\sigma}_{\nu}^{-}\right)_{+}
\equiv\hat{\sigma}_{\mu}^{+}\hat{\sigma}_{\nu}^{-}
+\hat{\sigma}_{\mu}^{-}\hat{\sigma}_{\nu}^{+}$.
As expected, the local heat current is only related to the dynamics of populations of the
TLSs, whereas the non-local heat current relies on the coherence of two TLSs in the same bath.

Analogously, for the same setting and explicit form of $\hat{V}_{i,n}$ of Eq.~(\ref{Vi}),
we obtain expressions for the local and non-local work currents as
\begin{eqnarray}\label{workcurr2loc}
\dot{W}^{com}_{loc}&=&\frac{1}{2}\sum_{i=h,c}\sum_{n=1}^{N}
\left\{\gamma_{i,nn}^{-}\left\langle\hat{\sigma}_{i,n}^{+}\hat{F}_{i,n}\right\rangle_{\rho_{S}}
-\gamma_{i,nn}^{+}\left\langle \hat{F}_{i,n}\hat{\sigma}_{i,n}^{+}\right\rangle_{\rho_{S}}\right\}
\nonumber\\
&&+\mathrm{c.c.},
\end{eqnarray}
and
\begin{eqnarray}\label{workcurr2nloc}
\dot{W}^{com}_{non-loc}&=&\frac{1}{2}\sum_{i=h,c}\sum_{n\neq n'=1}^{N}
\left\{\gamma_{i,nn'}^{-}\left\langle\hat{\sigma}_{i,n}^{+}\hat{F}_{i,n'}\right\rangle_{\rho_{S}}
-\gamma_{i,nn'}^{+}\left\langle \hat{F}_{i,n'}\hat{\sigma}_{i,n}^{+}\right\rangle_{\rho_{S}}\right\}\nonumber\\
&&+\mathrm{c.c.},
\end{eqnarray}
where $\hat{F}_{i,n}=\left[\hat{H}_{I},\hat{\sigma}_{i,n}^{-}\right]$ and ``c.c.'' denotes complex conjugate.
It is worth noting that the operator $\hat{F}_{i,n}$
connects the non-commutativity of jump operators of the TLSs
to the interaction Hamiltonian $H_I$ of TLSs across the ensembles.
It also indicates that
the types of $\hat{H}_{I}$ are responsible for the magnitudes of work,
which would be an important control factor used to modify the performance of QTMs.
However, since we do not know the specific form of $H_I$,
the relationship between the work current and the state of the system
remains uncertain at this point. In what follows,
we present two types of $H_I$ to clarify how
the system coherence determine work current.

\subsubsection{Effects of intrasystem interactions on the work for two pairs of TLSs}\label{IIIA2}
As discussed above, the intrasystem interactions play a crucial role
on the work, which can be used as a control factor to enhance the performance of QTMs.
Moreover, the interaction types between subsystems across the two ensembles in our setup are rich,
whose effects on the work deserve a comprehensive study.
To illustrate this issue, we consider two types of intrasystem interactions
and take $N=2$ as an example, where two TLSs $S_{h,1}$ and $S_{h,2}$ ($S_{c,1}$ and $S_{c,2}$)
are simultaneously coupled to the bath $E_h$ ($E_c$).
The first scenario [see the left panel of Fig.~\ref{MD}(c)] is that each TLS in one group interacts simultaneously
with all the TLSs in another group with the
interaction Hamiltonian given as
\begin{equation}\label{HI1}
\hat{H}_{I}^{(1)}=\sum_{n,n'=1}^{2}\Omega_{nn'}
\left(\hat{\sigma}_{h,n}^{+}\hat{\sigma}_{c,n'}^{-}+\hat{\sigma}_{h,n}^{-}\hat{\sigma}_{c,n'}^{+}\right)
\end{equation}
with $\Omega_{nn'}$ the coupling strength.
By introducing the collective operators $\hat{S}_{i}^{\pm}=
\hat{S}_{i}^{x}\pm i\hat{S}_{i}^{y}$ with
$\hat{S}_{i}^{\alpha}=\sum_{n=1}^{N}\hat{\sigma}_{i,n}^{\alpha}$ (with $\alpha=x,y,z$)
for the $N$ TLSs in the ensemble $S^{(i)}$ and assuming $\Omega_{nn'}=\Omega$,
$\hat{H}_{I}^{(1)}$ (\ref{HI1}) can be
rewritten as
\begin{equation}\label{HI1Coll}
\hat{H}_{I}^{(1)}=\Omega
\left(\hat{S}_{h}^{+}\hat{S}_{c}^{-}+\hat{S}_{h}^{-}\hat{S}_{c}^{+}\right).
\end{equation}
The interaction presented in Eq.~(\ref{HI1Coll}) suggests that each group of TLSs behaves
as a unit and simultaneously exchanges energy between them.
The second scenario [see the right panel of Fig.~\ref{MD}(c)] is that the $n-$th TLS in one ensemble
only interacts with the $n-$th TLS in the other one
with the Hamiltonian
\begin{equation}\label{HI2}
\hat{H}_{I}^{(2)}=\sum_{n=1}^{2}\Omega_{n}
\left(\hat{\sigma}_{h,n}^{+}\hat{\sigma}_{c,n}^{-}+\hat{\sigma}_{h,n}^{-}\hat{\sigma}_{c,n}^{+}\right),
\end{equation}
with $\Omega_{n}$ denoting the interacting strength.

For the first type of inner interactions, the local work current
can be derived as
\begin{eqnarray}\label{H1Wloc}
\dot{W}^{com (1)}_{loc}
&=&-\frac{1}{2}\left[\Omega_{11}\Gamma_{h1c1}
\left\langle\left(\hat{\sigma}_{h,1}^{+}\hat{\sigma}_{c,1}^{-}\right)_{+}\right\rangle_{\rho_{S}}
+\Omega_{22}\Gamma_{h2c2}\right.\nonumber\\
&&\times\left\langle\left(\hat{\sigma}_{h,2}^{+}\hat{\sigma}_{c,2}^{-}\right)_{+}\right\rangle_{\rho_{S}}
+\Omega_{12}\Gamma_{h1c2}
\left\langle\left(\hat{\sigma}_{h,1}^{+}\hat{\sigma}_{c,2}^{-}\right)_{+}\right\rangle_{\rho_{S}}\nonumber\\
&&\left.+\Omega_{21}\Gamma_{h2c1}
\left\langle\left(\hat{\sigma}_{h,2}^{+}\hat{\sigma}_{c,1}^{-}\right)_{+}\right\rangle_{\rho_{S}}\right],
\end{eqnarray}
where $\Gamma_{h1c1}=\gamma_{h,11}^{-}+\gamma_{h,11}^{+}+\gamma_{c,11}^{-}+\gamma_{c,11}^{+}$,
$\Gamma_{h2c2}=\gamma_{h,22}^{-}+\gamma_{h,22}^{+}+\gamma_{c,22}^{-}+\gamma_{c,22}^{+}$,
$\Gamma_{h1c2}=\gamma_{h,11}^{-}+\gamma_{h,11}^{+}+\gamma_{c,22}^{-}+\gamma_{c,22}^{+}$ and
$\Gamma_{h2c1}=\gamma_{h,22}^{-}+\gamma_{h,22}^{+}+\gamma_{c,11}^{-}+\gamma_{c,11}^{+}$.
The expression (\ref{H1Wloc}) indicates that the local work current $\dot{W}^{com(1)}_{loc}$
is closely related to the pairwise coherence of subsystems across the two ensembles
that have interactions, i.e.,
$S_{h,1}$-$S_{c,1}$, $S_{h,2}$-$S_{c,2}$, $S_{h,1}$-$S_{c,2}$ and $S_{h,2}$-$S_{c,1}$.
To clarify this point, we introduce the decoupled basis $\left\{\left|11\right\rangle_{S_{h,n}S_{c,n^{\prime}}},
\left|10\right\rangle_{S_{h,n}S_{c,n^{\prime}}},
\left|01\right\rangle_{S_{h,n}S_{c,n^{\prime}}},
\left|00\right\rangle_{S_{h,n}S_{c,n^{\prime}}}\right\}$
for these pairs of TLSs, with $\left|1\right\rangle$ $(\left|0\right\rangle)$
the excited (ground) state of a TLS.
In this way, we note that $\left\langle \hat{\sigma}_{h,n}^{+}\hat{\sigma}_{c,n^{\prime}}^{-}\right\rangle_{\rho_{S}}$
gives the coherence between the levels $\left|10\right\rangle_{S_{h,n}S_{c,n^{\prime}}}$
and $\left|01\right\rangle_{S_{h,n}S_{c,n^{\prime}}}$ of the reduced state of $S_{h,n}$
and $S_{c,n^{\prime}}$, and correspondingly
$\left\langle \left(\hat{\sigma}_{h,n}^{+}\hat{\sigma}_{c,n^{\prime}}^{-}\right)_{+}\right\rangle_{\rho_{S}}
=2\mathrm{Re}\left[\left\langle \hat{\sigma}_{h,n}^{+}\hat{\sigma}_{c,n^{\prime}}^{-}\right\rangle_{\rho_{S}}\right]$.
Therefore, the local work current $\dot{W}^{com(1)}_{loc}$ of the first type interaction
$\hat{H}_{I}^{(1)}$ is specifically determined by a coherence-related quantity
\begin{equation}\label{Ccomloc1}
\mathcal{C}^{com(1)}_{loc}=-2\sum_{n,n'=1}^{2}\mathrm{Re}\left[\left\langle \hat{\sigma}_{h,n}^{+}\hat{\sigma}_{c,n^{\prime}}^{-}\right\rangle_{\rho_{S}}\right].
\end{equation}

The non-local part of the work current for the first type of interactions
reads
\begin{eqnarray}\label{nonw1}
\dot{W}^{com(1)}_{non-loc}&=&\frac{1}{2}\left(\gamma_{h,12}^{-}-\gamma_{h,12}^{+}\right)
\left[\Omega_{11}\left\langle\hat{\sigma}_{h,1}^{z}\left(\hat{\sigma}_{h,2}^{+}\hat{\sigma}_{c,1}^{-}\right)_{+}
\right\rangle_{\rho_{S}}\right.\nonumber\\
&&+\Omega_{12}\left\langle\hat{\sigma}_{h,1}^{z}\left(\hat{\sigma}_{h,2}^{+}\hat{\sigma}_{c,2}^{-}
\right)_{+}\right\rangle_{\rho_{S}} +\Omega_{21}\left\langle\hat{\sigma}_{h,2}^{z}\left(\hat{\sigma}_{h,1}^{+}\hat{\sigma}_{c,1}^{-}
\right)_{+}\right\rangle_{\rho_{S}}\nonumber\\
&&\left.+\Omega_{22}\left\langle\hat{\sigma}_{h,2}^{z}
\left(\hat{\sigma}_{h,1}^{+}\hat{\sigma}_{c,2}^{-}\right)_{+}\right\rangle_{\rho_{S}}\right]\nonumber\\
&&
+\frac{1}{2}\left(\gamma_{c,12}^{-}-\gamma_{c,12}^{+}\right)
\left[\Omega_{11}\left\langle\hat{\sigma}_{c,1}^{z}\left(\hat{\sigma}_{c,2}^{+}\hat{\sigma}_{h,1}^{-}
\right)_{+}\right\rangle_{\rho_{S}}\right.\nonumber\\
&&
+\Omega_{21}\left\langle\hat{\sigma}_{c,1}^{z}\left(\hat{\sigma}_{c,2}^{+}\hat{\sigma}_{h,2}^{-}
\right)_{+}\right\rangle_{\rho_{S}}+\Omega_{12}\left\langle\hat{\sigma}_{c,2}^{z}\left(\hat{\sigma}_{c,1}^{+}\hat{\sigma}_{h,1}^{-}
\right)_{+}\right\rangle_{\rho_{S}}\nonumber \\
&&
+\left.\Omega_{22}\left\langle\hat{\sigma}_{c,2}^{z}\left(\hat{\sigma}_{c,1}^{+}\hat{\sigma}_{h,2}^{-}
\right)_{+}\right\rangle_{\rho_{S}}\right].
\end{eqnarray}
Based on the expression of Eq.~(\ref{nonw1}),
we observe that the non-local work current $\dot{W}^{com(1)}_{non-loc}$ is
intricately linked to the coherence exhibited by subsystems, both within the same bath and
spanning across two distinct baths.
This can be demonstrated by analyzing the terms that constitute $\dot{W}^{com(1)}_{non-loc}$.
For example, the coherence-related quantity in the first term, i.e., $\left\langle\hat{\sigma}_{h,1}^{z}\left(\hat{\sigma}_{h,2}^{+}\hat{\sigma}_{c,1}^{-}\right)_{+}
\right\rangle_{\rho_{S}}$, is contributed by the coherence of $S_{h,1}$,
$S_{h,2}$ and $S_{c,1}$ with the former two being in the same bath
and the last one in the other one. Note that the term $\left\langle\hat{\sigma}_{h,1}^{z}\left(\hat{\sigma}_{h,2}^{+}\hat{\sigma}_{c,1}^{-}\right)_{+}
\right\rangle_{\rho_{S}}$ cannot be decomposed into a direct product of
$\left\langle\hat{\sigma}_{h,1}^{z}\right\rangle_{\rho_{S}}$
and $\left\langle\left(\hat{\sigma}_{h,2}^{+}\hat{\sigma}_{c,1}^{-}\right)_{+}
\right\rangle_{\rho_{S}}$.
Here, we also define a coherence-related quantity
to determine the non-local work current as
\begin{equation}\label{Cnloccom1}
\mathcal{C}^{com(1)}_{non-loc}=\sum_{i\neq i^{\prime}=h,c }\sum_{n\neq \bar{n},n^{\prime}=1}^{2}
\left(\left\langle \hat{\sigma}_{i,n}^{z} \left(\hat{\sigma}_{i,\bar{n}}^{+}\hat{\sigma}_{i^{\prime},n^{\prime}}^{-}\right)_{+}
\right\rangle_{\rho_{S}}\right).
\end{equation}

Next, we deal with the situation where the intrasystem interactions is given by
$\hat{H}_{I}^{(2)}$ of Eq.~(\ref{HI2}).
In this case, the local component of work current is obtained as
\begin{eqnarray}\label{W2loc}
\dot{W}^{com(2)}_{loc}
&=&-\frac{1}{2}\left[\Omega_{1}\Gamma_{h1c1}
\left\langle\left(\hat{\sigma}_{h,1}^{+}\hat{\sigma}_{c,1}^{-}\right)_{+}\right\rangle_{\rho_{S}}\right.\nonumber\\
&&\left.+\Omega_{2}\Gamma_{h2c2}
\left\langle\left(\hat{\sigma}_{h,2}^{+}\hat{\sigma}_{c,2}^{-}\right)_{+}\right\rangle_{\rho_{S}}\right].
\end{eqnarray}
The above equation shows that $\dot{W}^{com(2)}_{loc}$
is related to the coherences of TLS pairs $S_{h,1}-S_{c,1}$ and $S_{h,2}-S_{c,2}$ that
have interactions and determined specifically by
\begin{equation}\label{Cloccom2}
\mathcal{C}^{com(2)}_{loc}=-2\sum_{n=1}^{2}\mathrm{Re}\left[\left\langle \hat{\sigma}_{h,n}^{+}\hat{\sigma}_{c,n}^{-}\right\rangle\right].
\end{equation}
It is confirmed once again that the local work current
in the simultaneous system-bath interaction is linked to inner interactions between subsystems
across the two baths.
Therefore, it is possible to control the work current and consequently achieve
different QTMs by changing the types of intrasystem interactions,
e.g., $\hat{H}_{I}^{(1)}$ (\ref{HI1})
and $\hat{H}_{I}^{(2)}$ (\ref{HI2}) we consider here.
Similarly, the non-local work current can be obtained as
\begin{eqnarray}\label{nonw2}
\dot{W}^{com(2)}_{non-loc}&=&\frac{1}{2}\left(\gamma_{h,12}^{-}-\gamma_{h,12}^{+}\right)
\left[\Omega_{1}\left\langle\hat{\sigma}_{h,1}^{z}\left(\hat{\sigma}_{h,2}^{+}\hat{\sigma}_{c,1}^{-}\right)_{+}
\right\rangle_{\rho_{S}}\right.\nonumber\\
&&\left.+\Omega_{2}\left\langle\hat{\sigma}_{h,2}^{z}
\left(\hat{\sigma}_{h,1}^{+}\hat{\sigma}_{c,2}^{-}\right)_{+}\right\rangle_{\rho_{S}}\right]\nonumber\\
&&
+\frac{1}{2}\left(\gamma_{c,12}^{-}-\gamma_{c,12}^{+}\right)
\left[\Omega_{1}\left\langle\hat{\sigma}_{c,1}^{z}\left(\hat{\sigma}_{c,2}^{+}\hat{\sigma}_{h,1}^{-}
\right)_{+}\right\rangle_{\rho_{S}}\right.\nonumber\\
&&\left.+\Omega_{2}\left\langle\hat{\sigma}_{c,2}^{z}\left(\hat{\sigma}_{c,1}^{+}\hat{\sigma}_{h,2}^{-}
\right)_{+}\right\rangle_{\rho_{S}}\right].
\end{eqnarray}
The expression of $\dot{W}^{com(2)}_{non-loc}$ clearly indicates that
the non-local work current is related to the coherence of
three subsystems with two in the same bath and the other one in the other bath.
A coherence-related quantity that
determines the non-local work current in this case can be defined as
\begin{equation}\label{Cnloccom2}
\mathcal{C}^{com(2)}_{non-loc}=\sum_{i\neq i^{\prime}=h,c }\sum_{n\neq \bar{n}=1}^{2}
\left(\left\langle \hat{\sigma}_{i,n}^{z} \left(\hat{\sigma}_{i,\bar{n}}^{+}\hat{\sigma}_{i^{\prime},n}^{-}\right)_{+}
\right\rangle_{\rho_{S}}\right).
\end{equation}

\subsection{Cascaded system-bath interactions}\label{IIIB}
Although both the cascaded and simultaneous system-bath
interactions can lead to non-local dissipation,
the differences of thermodynamic quantities in these two cases remain unclear.
It turns out that the work current of the cascaded model can also be formulated as
a sum of local and non-local components, i.e.,
$\dot{W}^{\mathrm{cas}}=\dot{W}^{\mathrm{cas}}_{loc}+\dot{W}^{\mathrm{cas}}_{non-loc}$.
Moreover, the local part $\dot{W}^{\mathrm{cas}}_{loc}$
has the same structure as the one given in Eq.~(\ref{wcurrloc}) for the common baths.
The difference of work current between the cascaded and common models
is evident in the non-local part, which can be derived as
\begin{equation}
\dot{W}^{\mathrm{cas}}_{non-loc}=-\sum_{i=h,c}\sum_{n^{\prime}>n=1}^{N}
\left\langle\left[\hat{V}_{i,n},\left[\hat{V}_{i,n'},
\hat{H}_{S}+\hat{H}_{E_{i}}\right]\right]\right\rangle_{\rho_{SE}}.
\end{equation}
In the same way, the heat current regarding bath $E^{(i)}$ can also be decomposed into
local and non-local ones, i.e.,
$\dot{Q}^{\mathrm{cas}(i)}=\dot{Q}^{\mathrm{cas}(i)}_{loc}+\dot{Q}^{\mathrm{cas}(i)}_{non-loc}$,
the former being consistent with that presented in Eq.~(\ref{Qloc}),
while the latter takes the form as
\begin{equation}
\dot{Q}^{\mathrm{cas}(i)}_{non-loc}=\sum_{n^{\prime}>n=1}^{N}\left\langle\left[\hat{V}_{i,n},\left[\hat{V}_{i,n'},\hat{H}_{E_{i}}\right]\right]\right\rangle_{\rho_{SE}}.
\end{equation}

For the specific case where the system and baths are composed of TLSs and harmonic oscillators,
respectively, we can obtain more concrete expressions for the
work and heat currents. As their local parts are the same as that for the simultaneous system-bath interaction model,
as given in Eqs.~(\ref{workcurr2loc}) and (\ref{Qinonloc}),
we only provide the non-local components being of the forms
\begin{eqnarray}\label{Wcasnon}
\dot{W}^{cas}_{non-loc}&=&\sum_{i=h,c}\sum_{n^{\prime}>n=1}^{N}\left\{\gamma_{i,nn'}^{-}\left\langle\hat{\sigma}_{i,n}^{+}\hat{F}_{i,n'}\right\rangle_{\rho_{S}}
-\gamma_{i,nn'}^{+}\left\langle \hat{F}_{i,n'}\hat{\sigma}_{i,n}^{+}\right\rangle_{\rho_{S}}\right\}\nonumber\\
&+&\mathrm{c.c},
\end{eqnarray}
with $\hat{F}_{i,n}=\left[\hat{H}_{I},\hat{\sigma}_{i,n}^{-}\right]$,
and
\begin{equation}
\dot{Q}^{cas(i)}_{non-loc}=\omega_{i}\sum_{n'>n=1}^{N}\left(\gamma_{i,nn'}^{+}
-\gamma_{i,nn'}^{-}\right)\left\langle\left(\hat{\sigma}_{i,n}^{+}\hat{\sigma}_{i,n'}^{-}\right)_{+}
\right\rangle_{\rho_{S}}.
\end{equation}
Obviously, the non-local heat current $\dot{Q}^{cas(i)}_{non-loc}$
is determined by the coherence of subsystems in the same bath,
being consistent to the results summarized in the Table \ref{table1}.
Regarding the non-local heat current, we first need to present the form of $H_I$,
which is discussed in detail subsequently.

\begin{figure*}[t!]
	\begin{center}
		{\includegraphics[width=0.35\linewidth]{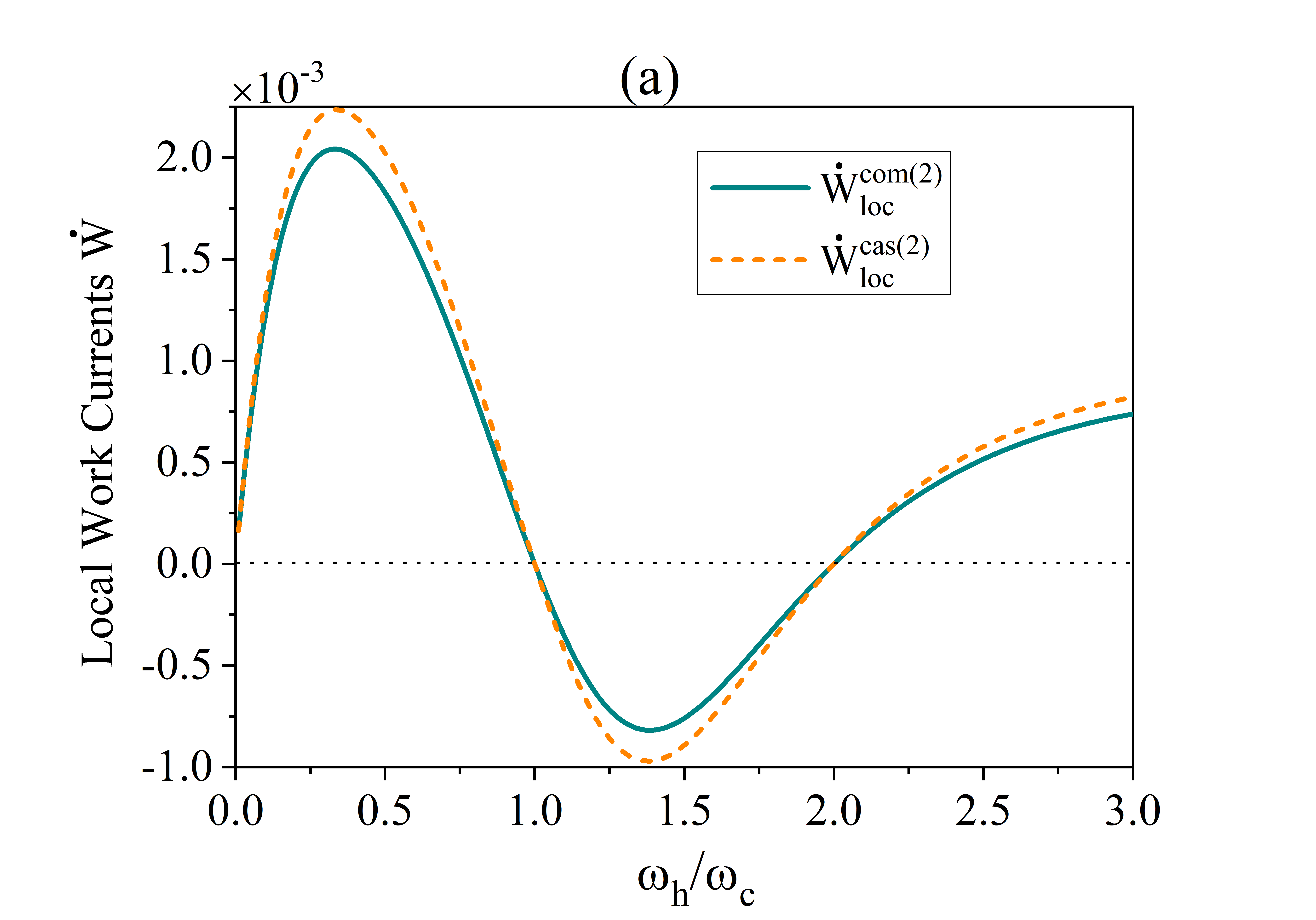} }
		{\includegraphics[width=0.35\linewidth]{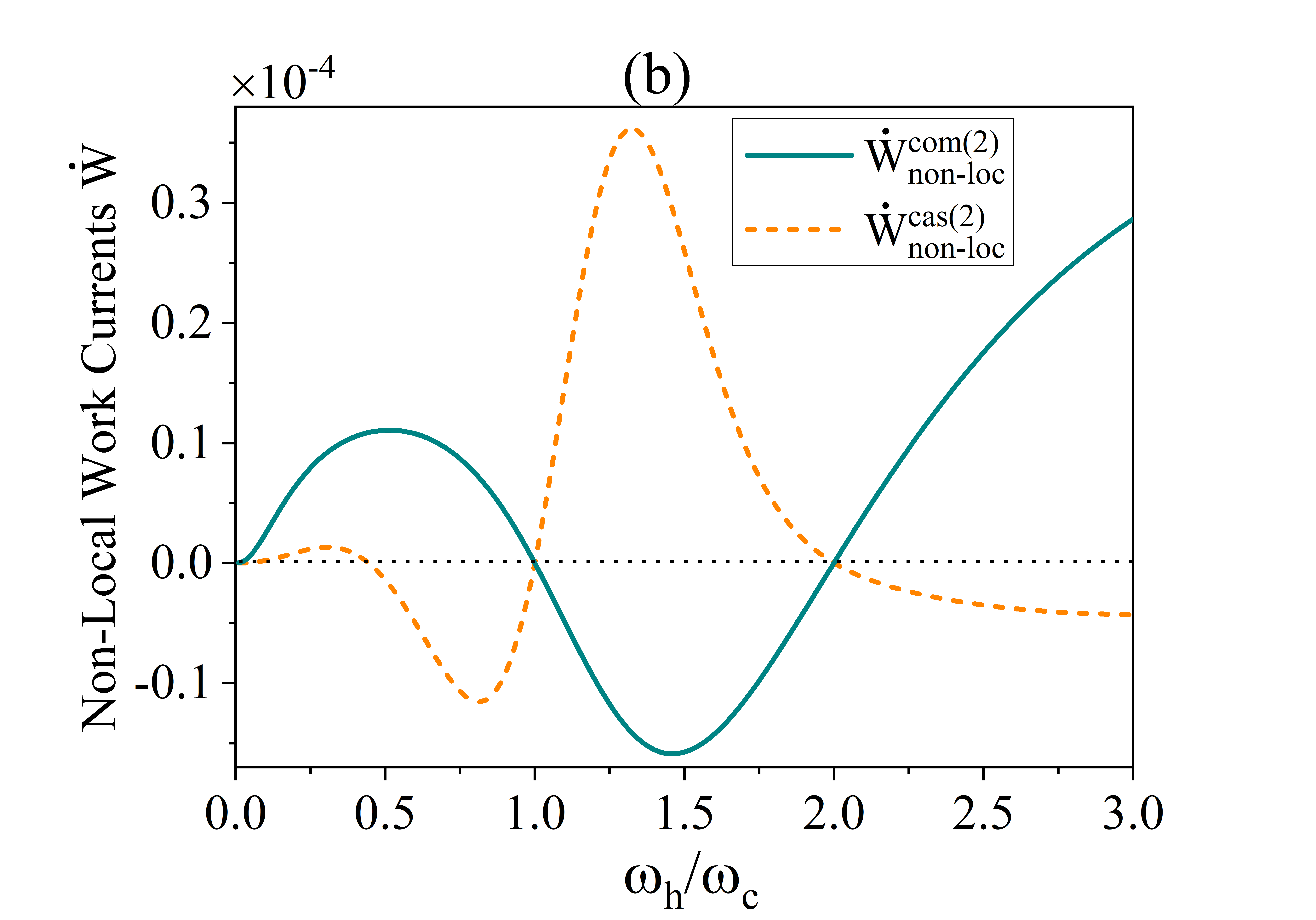} }
		{\includegraphics[width=0.35\linewidth]{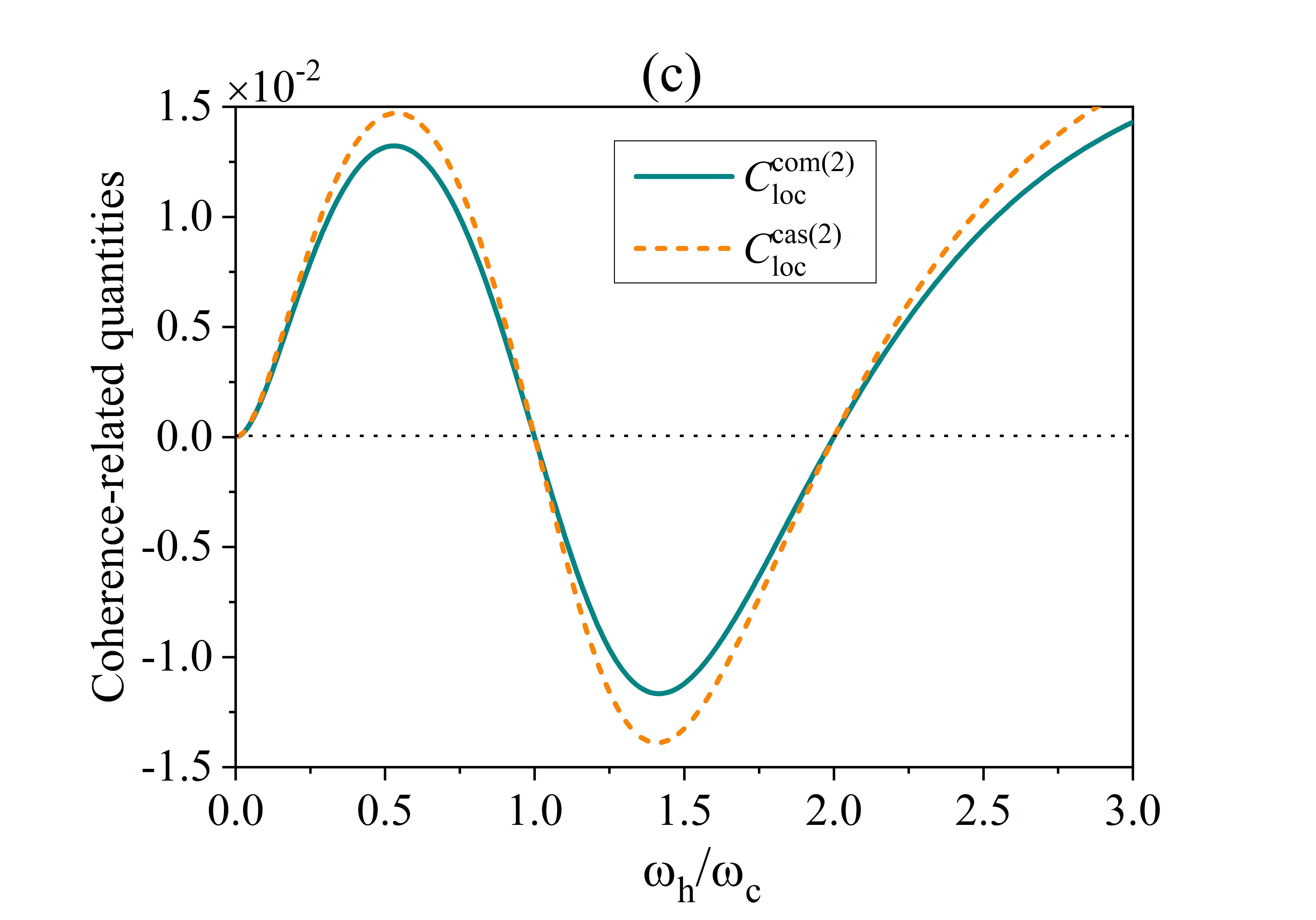} }
		{\includegraphics[width=0.35\linewidth]{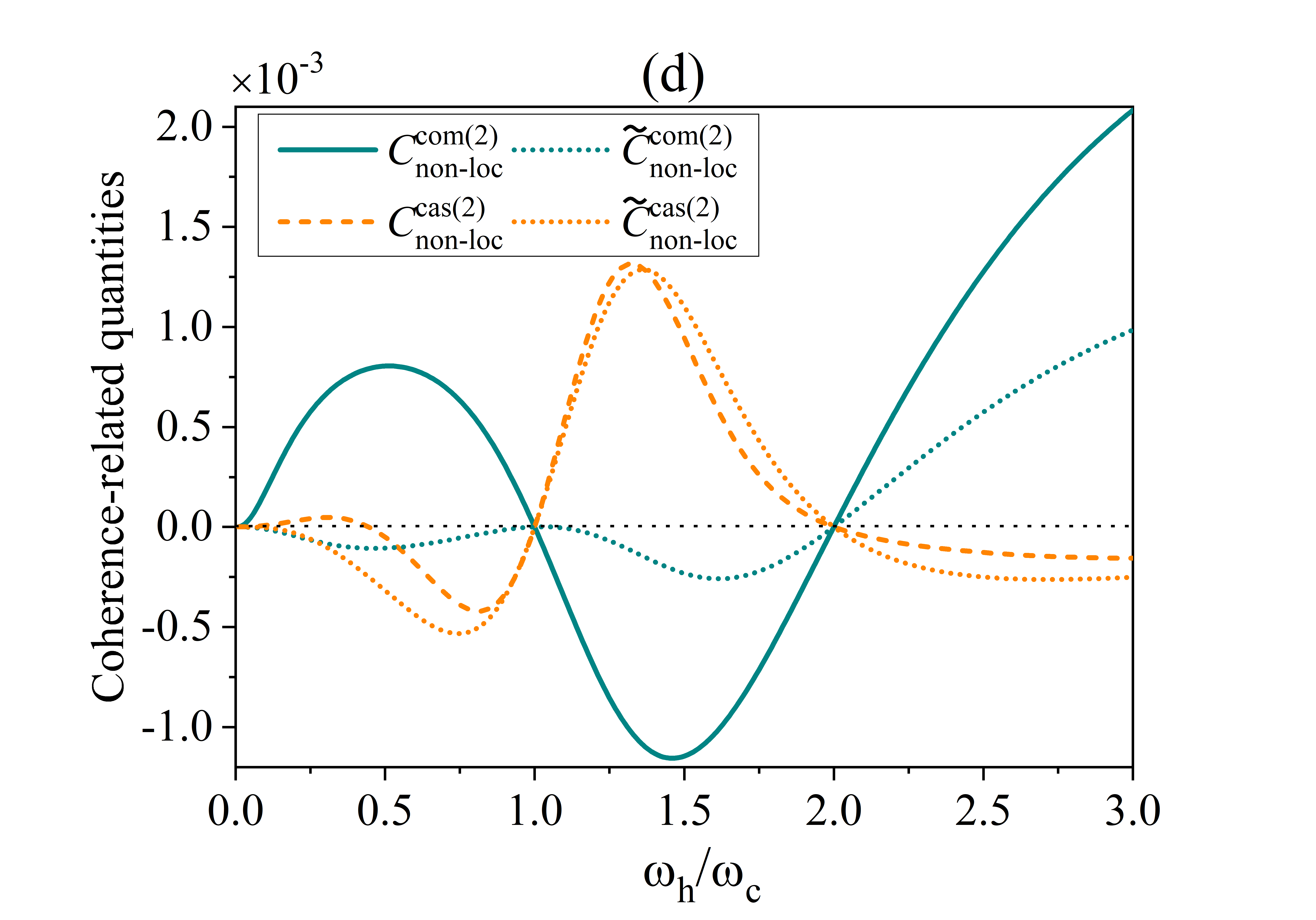} }
	\end{center}
	\caption{(a) The local components of work currents, i.e., $\dot{W}_{loc}^{com(2)}$
		and $\dot{W}_{loc}^{cas(2)}$, under the simultaneous and cascaded
		system-bath interactions for the second type of intrasystem couplings depicted by
		$\hat{H}_{I}^{(2)}$, Eq.~(\ref{HI2}), as a function of $\omega_{h}/\omega_{c}$.
		(b) The same as (a) but for the non-local work currents $\dot{W}_{non-loc}^{com(2)}$
		and $\dot{W}_{non-loc}^{cas(2)}$.
		(c) The coherence-related quantities $\mathcal{C}_{loc}^{com(2)}$ and $\mathcal{C}_{loc}^{cas(2)}$
		that determines the local work currents $\dot{W}_{loc}^{com(2)}$
		and $\dot{W}_{loc}^{cas(2)}$, respectively, against $\omega_{h}/\omega_{c}$.
		(d) The same as (c) but for $\mathcal{C}_{non-loc}^{com(2)}$ and $\mathcal{C}_{non-loc}^{cas(2)}$
		that determines the non-local work currents $\dot{W}_{non-loc}^{com(2)}$
		and $\dot{W}_{non-loc}^{cas(2)}$, respectively. Additionally, we includes the curves
		regarding $\mathcal{\widetilde{C}}_{non-loc}^{com(2)}$ and $\mathcal{\widetilde{C}}_{non-loc}^{cas(2)}$,
		given in Eqs.~(\ref{key1}) and (\ref{key2}), $\mathcal{C}_{non-loc}^{com(2)}\neq\mathcal{\widetilde{C}}_{non-loc}^{com(2)}$ and
		$\mathcal{C}_{non-loc}^{cas(2)}\neq\mathcal{\widetilde{C}}_{non-loc}^{cas(2)}$.
		The remaining parameters are set as $g_{h,1}=g_{c,1}=0.5\omega_{c}$, $g_{h,2}=g_{c,2}=0.55\omega_{c}$,
		$\Omega_{1}=\Omega_{2}=0.1\omega_{c}$, $T_{h}=2\omega_{c}$, and $T_{c}=\omega_{c}$.
	}	\label{coh}
\end{figure*}

For the cascaded model,
we also examine influences of the types of
intrasystem interactions, i.e., $\hat{H}_{I}^{(1)}$ in Eq.~(\ref{HI1}) and
$\hat{H}_{I}^{(2)}$ in Eq.~(\ref{HI2}), on the work currents.
The expressions of local work currents $\dot{W}^{cas(1)}_{loc}$ and $\dot{W}^{cas(2)}_{loc}$
associated with $\hat{H}_{I}^{(1)}$
and $\hat{H}_{I}^{(2)}$ are the same as that given in Eqs.~(\ref{H1Wloc}) and (\ref{W2loc})
for the common baths, respectively.
The coherence-related quantities $\mathcal{C}^{cas(1)}_{loc}$
and $\mathcal{C}^{cas(2)}_{loc}$ that determine the work currents
$\dot{W}^{cas(1)}_{loc}$ and $\dot{W}^{cas(2)}_{loc}$ are also possess identical
forms as that given in Eqs.~(\ref{Ccomloc1}) and (\ref{Cloccom2}).
The differences in work currents of the cascaded model
compared to the simultaneous interaction model are embodied in their non-local parts,
which can be formulated with respect to $\hat{H}_{I}^{(1)}$ and $\hat{H}_{I}^{(2)}$ as
\begin{eqnarray}\label{Wcas1}
\dot{W}^{cas(1)}_{non-loc}&=&\left(\gamma_{h,12}^{-}-\gamma_{h,12}^{+}\right)
\left[\Omega_{21}\left\langle\hat{\sigma}_{h,2}^{z}\left(\hat{\sigma}_{h,1}^{+}\hat{\sigma}_{c,1}^{-}\right)_{+}\right\rangle_{\rho_{S}}\right.\nonumber\\
&+&\left.\Omega_{22}\left\langle\hat{\sigma}_{h,2}^{z}\left(\hat{\sigma}_{h,1}^{+}\hat{\sigma}_{c,2}^{-}\right)_{+}\right\rangle_{\rho_{S}}\right]\nonumber\\
&+&\left(\gamma_{c,12}^{-}-\gamma_{c,12}^{+}\right)
\left[\Omega_{12}\left\langle\hat{\sigma}_{c,2}^{z}\left(\hat{\sigma}_{c,1}^{+}\hat{\sigma}_{h,1}^{-}\right)_{+}\right\rangle_{\rho_{S}}\right.\nonumber\\
&+&\left.\Omega_{22}\left\langle\hat{\sigma}_{c,2}^{z}\left(\hat{\sigma}_{c,1}^{+}\hat{\sigma}_{h,2}^{-}\right)_{+}\right\rangle_{\rho_{S}}\right],
\end{eqnarray}
and
\begin{eqnarray}\label{Wcas2}
\dot{W}^{cas(2)}_{non-loc}&=&\Omega_{2}\left(\gamma_{h,12}^{-}-\gamma_{h,12}^{+}\right)
\left[\left\langle\hat{\sigma}_{h,2}^{z}
\left(\hat{\sigma}_{h,1}^{+}\hat{\sigma}_{c,2}^{-}\right)_{+}\right\rangle_{\rho_{S}}\right]\nonumber\\
&+&\Omega_{2}\left(\gamma_{c,12}^{-}-\gamma_{c,12}^{+}\right)
\left[\left\langle\hat{\sigma}_{c,2}^{z}\left(\hat{\sigma}_{c,1}^{+}\hat{\sigma}_{h,2}^{-}
\right)_{+}\right\rangle_{\rho_{S}}\right],
\end{eqnarray}
respectively. It is apparent that the non-local work current in the cascaded model is influenced by the coherence of three subsystems: those within the same bath as well as spanning across the two baths.
The coherence-related quantities that
determine $\mathcal{C}^{cas(1)}_{non-loc}$ and $\mathcal{C}^{cas(2)}_{non-loc}$
can also be given as
\begin{equation}\label{Cnloccas1}
\mathcal{C}^{cas(1)}_{non-loc}=\sum_{i\neq i^{\prime}=h,c }\sum_{n=1}^{2}
\left(\left\langle \hat{\sigma}_{i,2}^{z} \left(\hat{\sigma}_{i,1}^{+}\hat{\sigma}_{i^{\prime},n}^{-}\right)_{+}
\right\rangle_{\rho_{S}}\right),
\end{equation}
and
\begin{equation}\label{Cnloccas2}
\mathcal{C}^{cas(2)}_{non-loc}=\sum_{i\neq i^{\prime}=h,c }
\left(\left\langle \hat{\sigma}_{i,2}^{z} \left(\hat{\sigma}_{i,1}^{+}\hat{\sigma}_{i^{\prime},2}^{-}\right)_{+}
\right\rangle_{\rho_{S}}\right).
\end{equation}

According to the cascaded model, the prior interaction of $S_{h,1}$ ($S_{c,1}$) with
the corresponding bath has a unidirectional impact on the subsequent
interaction of $S_{h,2}$ ($S_{c,2}$) with the bath.
Therefore, although the correlation between $S_{h,1}$ and $S_{h,2}$ ($S_{c,1}$ and $S_{c,2}$)
can be generated by the non-local dissipation, its influence is manifested only in the dynamics of $S_{h,2}$ ($S_{c,2}$).
As shown by Eqs.~(\ref{Wcas1}) and (\ref{Wcas2}), this unique one-way effect of the cascaded model is incorporated into the non-local work currents, presenting a notable difference compared to the simultaneous system-bath interaction scenario.
This is evident by comparing the constituent terms of non-local work currents
given in Eqs.~(\ref{Wcas1}) and (\ref{Wcas2})
with those in Eqs.~(\ref{nonw1}) and (\ref{nonw2})].

\subsection{Evidence of relationships between work currents and coherence-related quantities}\label{IIIC}
In this section, we illustrate the relationships between work currents, including both local and non-local
components, and the associated coherence-related quantities.
We still focus on a system with $N=2$, in which the TLSs $S_{h,1}$ and $S_{h,2}$ are coupled to the hot
bath $E_h$, while $S_{c,1}$ and $S_{c,2}$ are coupled to
the cold one $E_c$. The bath ancillas are modeled as harmonic oscillators.
For both common and cascaded models, we consider the second type of interaction within the system,
Eq.~(\ref{HI2}), which involves only the couplings of $S_{h,1}$-$S_{c,1}$ and $S_{h,2}$-$S_{c,2}$.
It can allow us to specify more clearly the dependence of non-local work current
on the coherence of TLSs in the same bath and without direct interactions.

In Fig.~\ref{coh}(a), we depict the variations of local work currents
$\dot{W}_{loc}^{com(2)}$ and $\dot{W}_{loc}^{cas(2)}$ under the simultaneous and cascaded
interaction models
as a function of $\omega_{h}/\omega_{c}$.
We observe that the magnitudes of work currents differ in these two models
and generally we have $|\dot{W}_{loc}^{cas(2)}|>|\dot{W}_{loc}^{com(2)}|$.
Fig.~\ref{coh}(c) illustrates the coherence-related quantities
$\mathcal{C}_{loc}^{com(2)}$ and $\mathcal{C}_{loc}^{cas(2)}$ which
govern the work currents $\dot{W}_{loc}^{com(2)}$ and $\dot{W}_{loc}^{cas(2)}$, respectively.
The changing trends of these coherence-related quantities coincides with
the work currents shown in Fig.~\ref{coh}(a), further indicating that the latter
are determined by the coherences of subsystems having interactions.

In Fig.~\ref{coh}(b), we illustrate the changes of non-local work currents
$\dot{W}_{non-loc}^{com(2)}$ and $\dot{W}_{non-loc}^{cas(2)}$
against $\omega_{h}/\omega_{c}$.
Unlike the local work currents shown in Fig.~\ref{coh}(a),
the directions of non-local work currents $\dot{W}_{non-loc}^{com(2)}$ and $\dot{W}_{non-loc}^{cas(2)}$
under different models can be opposite within the same intervals.
Fig.~\ref{coh}(d) displays the coherence-related quantities
$\mathcal{C}_{non-loc}^{com(2)}$ of Eq.~(\ref{Cnloccom2}) and
$\mathcal{C}_{non-loc}^{cas(2)}$ of Eq.~(\ref{Cnloccas2}), whose
behaviors are consistent with the non-local work currents shown in Fig.~\ref{coh}(b).
Additionally, we also include the curves of $\mathcal{\widetilde{C}}_{non-loc}^{com(2)}$
and $\mathcal{\widetilde{C}}_{non-loc}^{cas(2)}$ to demonstrate that the coherence-related quantities
cannot be decomposed into a product form in the sense of $\mathcal{C}_{non-loc}^{com(2)}\neq\mathcal{\widetilde{C}}_{non-loc}^{com(2)}$ and
$\mathcal{C}_{non-loc}^{cas(2)}\neq\mathcal{\widetilde{C}}_{non-loc}^{cas(2)}$
with
\begin{equation}\label{key1}
\mathcal{\widetilde{C}}_{non-loc}^{com(2)}=\sum_{i\neq i^{\prime}=h,c }\sum_{n\neq \bar{n}=1}^{2}
\left(\left\langle \hat{\sigma}_{i,n}^{z}\right\rangle_{\rho_{S}}\otimes\left\langle \left(\hat{\sigma}_{i,\bar{n}}^{+}\hat{\sigma}_{i^{\prime},n}^{-}\right)_{+}
\right\rangle_{\rho_{S}}\right)
\end{equation}
and
\begin{equation}\label{key2}
\mathcal{\widetilde{C}}_{non-loc}^{cas(2)}=\sum_{i\neq i^{\prime}=h,c }
\left(\left\langle \hat{\sigma}_{i,2}^{z}\right\rangle_{\rho_{S}}\otimes \left\langle \left(\hat{\sigma}_{i,1}^{+}\hat{\sigma}_{i^{\prime},2}^{-}\right)_{+}
\right\rangle_{\rho_{S}}\right).
\end{equation}

\section{Quantum thermal machines}\label{IV}
In this section, we address the possible uses of our scheme as QTMs and compare their performance under different situations.

To be specific, we consider the configuration
where each ensemble is composed of two TLSs and the bath
ancilla is the harmonic oscillator.
We compare the performance of QTMs for both simultaneous and cascaded system-bath interactions, entailing two different types of intrasystem interactions given in Eqs.~(\ref{HI1}) and (\ref{HI2}), respectively.
Moreover, as a benchmark for assessing the effect of non-local dissipation on thermal machine performance,
we take the case where each TLS in the ensemble independently interacts with the bath (i.e.,
the system undergoes independent dissipations),
which is obtained by removing the cross terms characterizing non-local dissipations in the QME of Eq.~(\ref{Dinonloc}).
We label the work and heat currents under independent dissipation as $\dot{W}^{ind(1)}$ ($\dot{W}^{ind(2)}$)
and $\dot{Q}_{i}^{ind(1)}$ ($\dot{Q}_{i}^{ind(2)}$) for the first (second) type
of intrasystem interactions, respectively.
As a result, there exist six different scenarios for the operations of QTMs which are to be compared.

\begin{figure}[t!]
	\begin{center}
		{\includegraphics[width=0.85\linewidth]{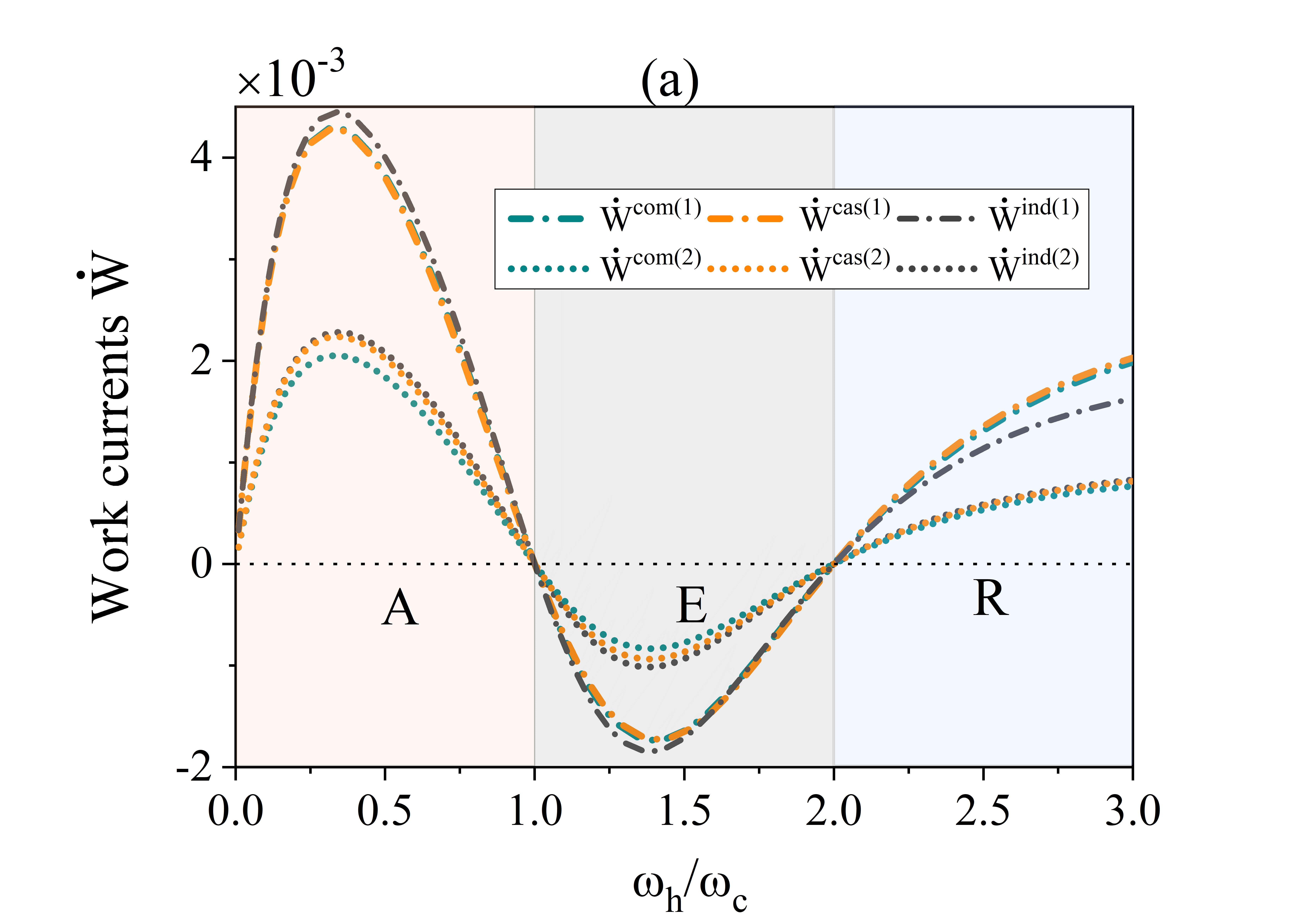} }
		{\includegraphics[width=0.85\linewidth]{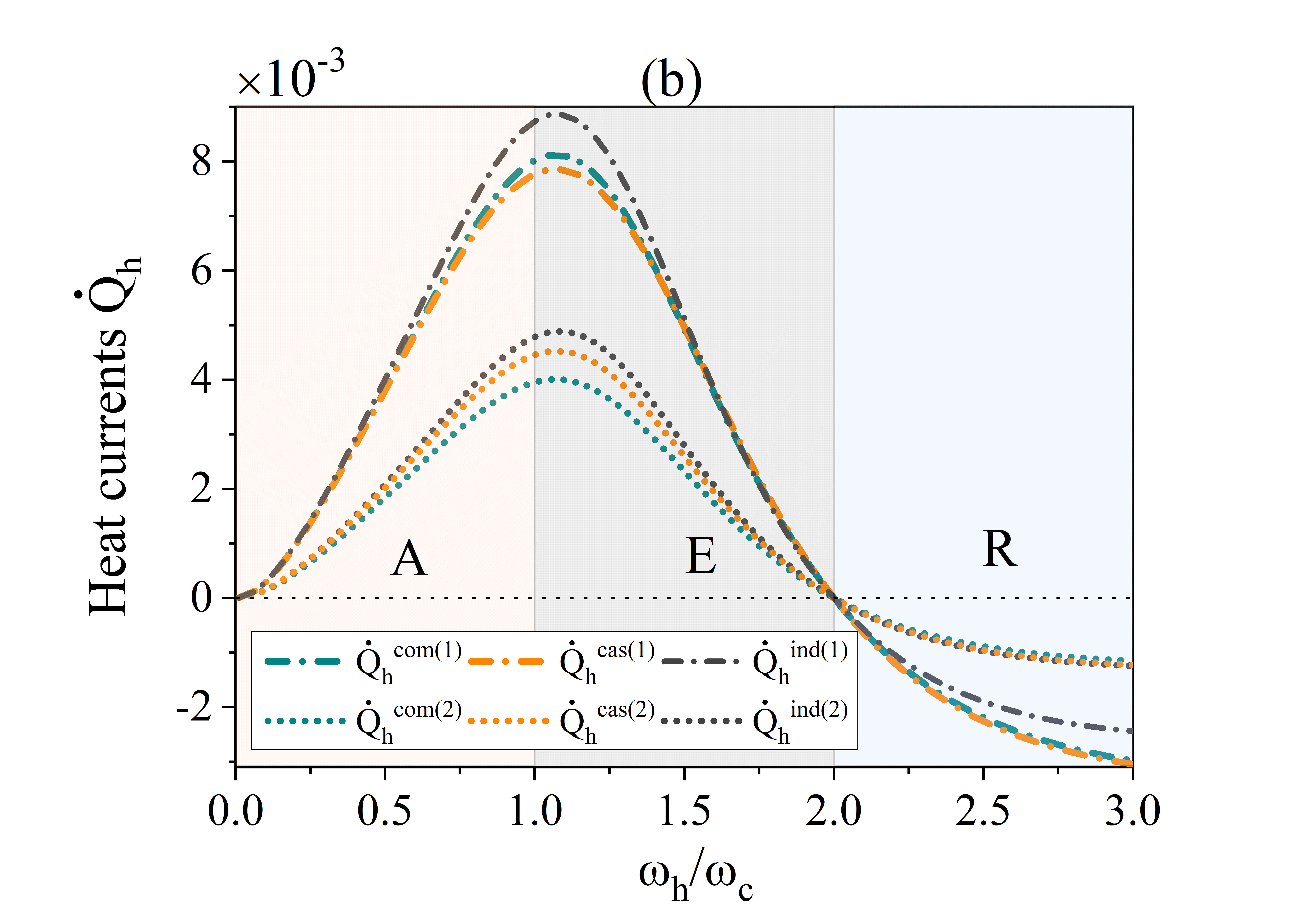} }
		{\includegraphics[width=0.85\linewidth]{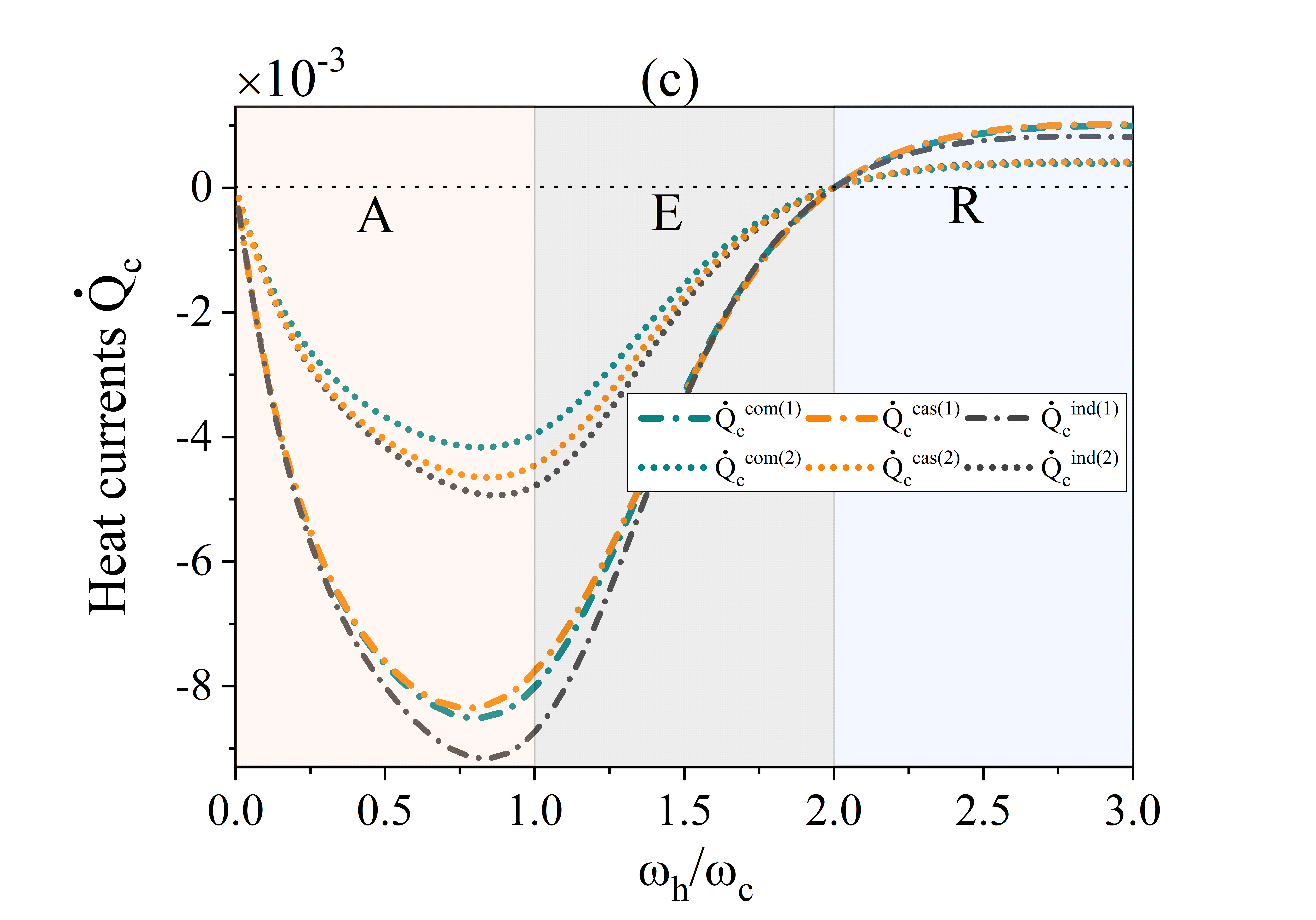} }
	\end{center}
	\caption{Work currents $\dot{W}$ (a), heat currents $\dot{Q}_{h}$
		from the hot bath (b), and $\dot{Q}_{c}$ from the cold bath (c), in the stationary regime,
		as a function of $\omega_{h}/\omega_{c}$. Both simultaneous and cascaded system-bath interactions are considered, with two different types of intrasystem interactions. Three operating regimes of the system of the quantum thermal machine are identified
		based on the signs of work and heat currents:
		accelerator (A), engine (E), and refrigerator (R).
		For the first type of intrasystem interaction we set $\Omega_{11}=\Omega_{12}=\Omega_{21}=\Omega_{22}=0.1\omega_{c}$.
		The other parameters are the same as those of Fig.~\ref{coh}. }	\label{machine}
\end{figure}

It turns out that the machine can achieve three operating regimes, namely refrigerator, engine, and accelerator (oven) depending on the choices of
$\omega_{h}/\omega_{c}$, as shown in Fig.~\ref{machine}.
By definition, the thermodynamic quantities
are positive when the energy enters the system, therefore the functions of QTMs
can be identified based on the directions of work and heat currents \cite{Buffoni2019,Solfanelli2020,Sur2023}. Therefore, 
the refrigerator (R) is characterized by $\dot{W}>0$, $\dot{Q}_{h}<0$ and
$\dot{Q}_{c}>0$; the engine (E) by $\dot{W}<0$, $\dot{Q}_{h}>0$ and $\dot{Q}_{c}<0$; the accelerator (A) by $\dot{W}>0$, $\dot{Q}_{h}>0$ and
$\dot{Q}_{c}<0$. The thermodynamic quantities
$\dot{W}$, $\dot{Q}_{h}$, and $\dot{Q}_{c}$ are plotted in 
panels (a), (b), and (c) of Fig.~\ref{machine}, respectively.
The combination of these three panels displays the intervals in which the three functions (A, E, R) appear.

\begin{figure}[t!]
	\begin{center}
		{\includegraphics[width=\linewidth]{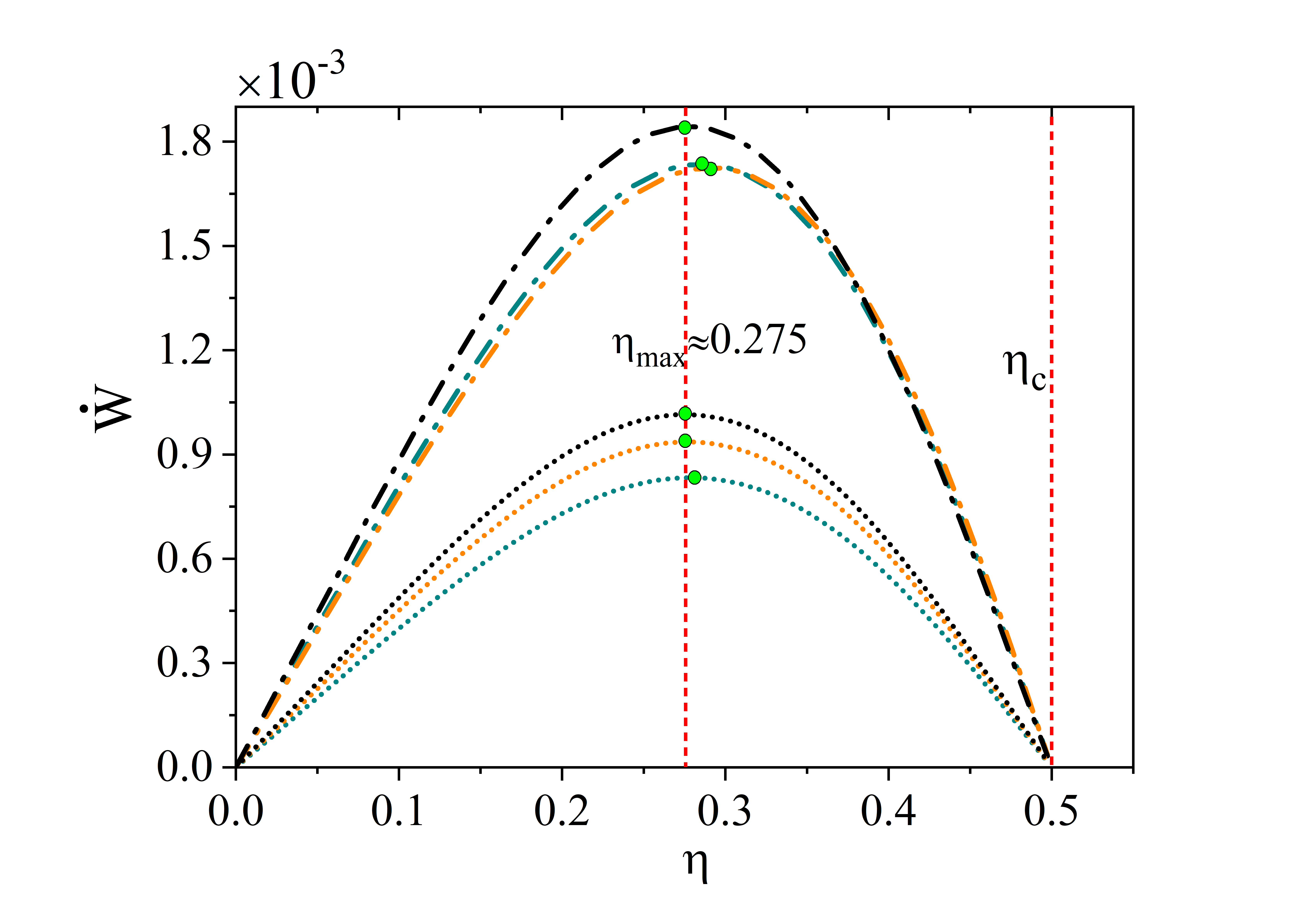} }
	\end{center}
	\caption{Parametric plot of power $\dot{W}$ versus efficiency $\eta$ of the engine
		for all six scenarios shown in Fig.~\ref{machine}(a) with the same styles for the various curves.
		The green dots mark efficiency at maximum power $\eta_{max}$. The left vertical dashed line corresponds to $\eta_{max}\approx0.275$ which can be attained
		by three cases, i.e., the first type of intrasystem interactions with independent baths,
		and the second type of intrasystem interactions with independent and cascaded baths.
		The right vertical dashed line corresponds to the Carnot efficiency $\eta_{c}=0.5$
		as the temperatures of the hot and cold baths are chosen as $T_{h}=2T_{c}$.
		The other parameters are the same as those of Fig.~\ref{machine}. }	\label{effi}
\end{figure}

We begin with a scenario where the setup functions as
an accelerator within the interval $0<\omega_{h}/\omega_{c}<1$,
based on our chosen parameters. The accelerator uses the external work injected
to enhance the heat flow from the hot bath to the cold bath.
The coefficient of performance (COP) of our accelerator is given as
$\mathrm{COP}_A=\dot{Q}_{h}/\dot{W}=\omega_{h}/\left(\omega_{h}-\omega_{c}\right)$,
which is only dependent on $\omega_{h}$ and $\omega_{c}$
and thus identical for all the considered cases.
In the presence of the same $COP_A$, the performance of the accelerator
can be characterized by the pumped heat $\dot{Q}_{h}$ from the hot
bath. We observe from Fig.~\ref{machine} (b) that $\dot{Q}_{h}$
for the first type of intrasystem interaction $\hat{H}_{I}^{(1)}$,
whether it is from the common bath model denoted as
$\dot{Q}_{h}^{\mathrm{com}(1)}$, the cascaded model as
$\dot{Q}_{h}^{\mathrm{cas}(1)}$, or the independent dissipation as $\dot{Q}_{h}^{\mathrm{ind}(1)}$,
is always larger than that for the second type of intrasystem interaction $\hat{H}_{I}^{(2)}$,
given as $\dot{Q}_{h}^{\mathrm{com}(2)}$, $\dot{Q}_{h}^{\mathrm{cas}(2)}$ and $\dot{Q}_{h}^{\mathrm{ind}(2)}$, respectively,
for these three situations.
Moreover, for both scenarios of $\hat{H}_{I}^{(1)}$ and $\hat{H}_{I}^{(2)}$,
the heat current $\dot{Q}_{h}$ under non-local dissipation, entailing
$\dot{Q}_{h}^{\mathrm{com}(1)}$ and $\dot{Q}_{h}^{\mathrm{cas}(1)}$ associated
with $\hat{H}_{I}^{(1)}$ and $\dot{Q}_{h}^{\mathrm{com}(2)}$ and $\dot{Q}_{h}^{\mathrm{cas}(2)}$
associated with $\hat{H}_{I}^{(2)}$, always
has a slightly smaller value than that of corresponding local dissipation,
i.e., $\dot{Q}_{h}^{\mathrm{ind}(1)}$ and $\dot{Q}_{h}^{\mathrm{ind}(2)}$.
This suggests that local dissipation is better than non-local dissipation.
However, local dissipation requires necessary space distances for individual subsystems, which
can be challenging in practice, especially for the first type of intrasystem interaction.
We can also see that whether simultaneous or cascaded system-bath interactions
are more advantageous for thermal machine performance is related to
the type of intrasystem interactions $\hat{H}_{I}^{(1)}$ and $\hat{H}_{I}^{(2)}$.
When considering $\hat{H}_{I}^{(1)}$,
the common bath shows slightly better performance with $\dot{Q}_{h}^{\mathrm{com}(1)}>\dot{Q}_{h}^{\mathrm{cas}(1)}$, whereas for $\hat{H}_{I}^{(2)}$, the cascaded system-bath interaction offers greater advantages
with $\dot{Q}_{h}^{\mathrm{cas}(2)}>\dot{Q}_{h}^{\mathrm{com}(2)}$.

By setting $1<\omega_{h}/\omega_{c}<2$, the thermal machine operates as an engine,
where the system absorbs heat from the hot bath
to perform work and pours the remaining heat into the cold bath.
The efficiency of the engine is given as $\eta=|\dot{W}|/\dot{Q}_{h}=1-\omega_{c}/\omega_{h}$, which is thus
independent of the considered situations given frequencies of the system.
To assess the performance of the thermal machine, we examine the extracted
work $|\dot{W}|$ under different scenarios, which can be observed from
Fig. \ref{machine} (a) within the interval
$1<\omega_{h}/\omega_{c}<2$. We note that most trends of $|\dot{W}|$
under various situations are similar
to the results obtained in the regime of accelerator.
We also notice a difference from interval $1.5<\omega_{h}/\omega_{c}<2$,
namely, in the case of the first type of intrasystem interaction,
independent dissipation may not necessarily be superior to non-local dissipation
as $|\dot{W}^{ind(1)}|$ is slightly smaller than $|\dot{W}^{cas(1)}|$.

At the point $\omega_{h}/\omega_{c}=1$, the work current becomes zero since the
energy conservation is satisfied and the same amount of heat is transferred from
the hot bath to the cold one. For the range of $\omega_{h}/\omega_{c}>2$,
the thermal machine functions as the refrigerator which transfers
heat from the cold bath to the hot one driven by the work injected through external source.
The COP of the refrigerator is found to be $\mathrm{COP}_R=\dot{Q}_{c}/\dot{W}=\omega_{c}/(\omega_{h}-\omega_{c})$,
which is also determined only by the frequencies of the system.
With identical COPs, the performance of the refrigerator is provided by the cooling power
$\dot{Q}_{c}$, which is shown in the interval $\omega_{h}/\omega_{c}>2$ of Fig.~\ref{machine} (c).
In contrast to previous discussions on accelerator and engine regimes,
the most significant difference is that non-local dissipation showcases advantages
over independent dissipation.
This is especially evident for the first type of intrasystem interaction
with both $\dot{Q}_{c}^{com(1)}$ and $\dot{Q}_{c}^{cas(1)}$ are larger than $\dot{Q}_{c}^{ind(1)}$.

Finally, we investigate the trade-off between the work current and efficiency of the engine
by constructing a parametric plot relative to these two quantities for all six scenarios,
as shown in Fig.~\ref{effi}. Interestingly, the efficiency
at maximum power $\eta_{max}$ (marked by green dots) varies depending on the specific
situations being considered. We see that
three scenarios have the same $\eta_{max}\approx0.275$,
namely, the first type of intrasystem interactions with independent baths,
and the second type of intrasystem interactions with independent and cascaded baths.
We also observe that, for the first type of intrasystem interactions,
$\eta_{max}$ under non-local dissipation (i.e., under the simultaneous and cascaded interaction models)
can be larger than that under independent dissipation. The work current vanishes
when approaching the Carnot efficiency $\eta_{c}=1-T_{c}/T_{h}$,
which corresponds to the reversible engine with zero entropy production.

\section{Conclusions}\label{V}

In this study, we have analyzed the effects
of steady-state coherence of a multipartite open quantum system
on relevant thermodynamic quantities, such as work and heat currents, and on quantum thermal machines (QTMs).
Our system consists of two many-particle ensembles,
each one coupled to a thermal bath. We have considered
that particles in an ensemble interact with the bath either simultaneously or sequentially,
named simultaneous and cascaded system-bath interaction models, respectively.
Both forms of interaction can lead to non-local dissipation in the system, which appears
not only in the derived master equation but also in the emergence of non-local heat and work currents.

Based on whether particles (subsystems) are in the same bath or different baths,
we classify the steady-state quantum coherence of the system into two categories:
coherence of particles in the same bath and coherence of particles spanning across two baths.
By definition, the particles in the former case do not have any mutual interactions,
whereas in the latter case they have direct interactions to each other.
We observe that, in addition to the local heat current being related to the populations
of the system, the non-local heat current as well as both the local and non-local work
currents are connected to the steady-state coherence of the system.
However, their dependence on the coherence between particles in the system is different:
the non-local heat current depends on the coherence of
particles in the same bath, the local work current
is due to the coherence of particles in different baths,
while the non-local work current is linked to the coherence of particles both in the same bath and in different baths. This result is summarized in Table \ref{table1}.

Furthermore, since the work current depends on the direct interactions between the two ensembles,
we have taken two types of intrasystem interactions into account.
For each scenario, we have provided the explicit expressions of
coherence-related quantities that determine the work currents.
We have also shown that, as a typical characteristic of the cascaded model,
the one-way influence between the dynamics of subsystems emerges
in the expressions of non-local work and heat currents through the coherence determining them.

Regarding QTMs, we have found that our scheme can function as an engine, a refrigerator, and an accelerator by suitably adjusting the transition frequency of the particles (two-level subsystems) in the two ensembles. We evaluate and compare the performance of QTMs across a total of six configurations, encompassing simultaneous, cascaded, and independent (local) system-bath interaction models, where each configuration presents two distinct types of intrasystem interactions. It turns out that the optimal performance scenario among the six configurations is
not fixed but depends both on the specific function of the QTM and on the selected parameter range. Consequently, independent dissipation and non-local dissipation, together with various types of intrasystem interactions, can be adjusted to effectively tailor the performance of diverse QTMs. 

Ultimately, our results demonstrate a link between thermodynamic quantities and quantum coherence, supplying useful insights for the design of quantum thermal machines. These findings thus motivate further studies about the control of quantum thermodynamic processes based on suitably engineered ensembles of quantum particles.

\acknowledgements
This work was supported by
National Natural Science Foundation (China) under Grant No. 12274257
and No. 11974209, Natural Science Foundation of Shandong Province (China) under Grant No. ZR2023LLZ015,
Taishan Scholar Project of Shandong Province (China) under Grant No. tsqn201812059,
and Youth Technological Innovation Support Program of Shandong Provincial Colleges and Universities under Grant No. 2019KJJ015. 
F.N. acknowledges support by the I+D+i project MADQuantum-CM, financed by the European Union NextGeneration-EU, Madrid Government and by the PRTR.
R.L.F. acknowledges support by MUR (Ministero dell’Università e della Ricerca) through the following projects: PNRR Project ICON-Q – Partenariato Esteso NQSTI – PE00000023 – Spoke 2 – CUP: J13C22000680006, PNRR Project QUANTIP – Partenariato Esteso NQSTI – PE00000023 – Spoke 9 – CUP: E63C22002180006. 

\vskip 0.1cm
\noindent\textbf{Data availability statement}
\vskip 0.1cm
\noindent The data cannot be made publicly available upon publication because they are not available in a format that
is sufficiently accessible or reusable by other researchers. The data that support the findings of this study are
available upon reasonable request from the authors.


\appendix

\section{Derivations of the general expressions of work and heat currents of Eqs.~(\ref{workcurr})-(\ref{Qnonloc})} \label{app:WQ}

During a collision, the change of work can be expressed as (see also Eq.~(\ref{DW})
in the main text)
\begin{equation}
\Delta W=\frac{1}{\sqrt{\tau}}\sum_{i=h,c}\sum_{n=1}^{N}\left[\left\langle \hat{V}_{i,n}\right\rangle_{\rho_{SE}}-\left\langle \hat{V}_{i,n}\right\rangle_{\rho'_{SE}}\right].
\end{equation}
The expectation value of $\hat{V}_{i,n}$ with respect the evolved state $\rho'_{SE}$ can be derived as follows,
\begin{eqnarray}
\left\langle \hat{V}_{i,n}\right\rangle_{\rho'_{SE}}&=&\left\langle\hat{U}\left(\tau\right)\hat{V}_{i,n}\hat{U}^{\dagger}\left(\tau\right)\right\rangle_{\rho_{SE}}\nonumber\\
&=&\left\langle e^{-i\tau\hat{H}_{tot}}\hat{V}_{i,n}e^{i\tau\hat{H}_{tot}}\right\rangle_{\rho_{SE}}\nonumber\\
&=&\left\langle\left(1-i\tau\hat{H}_{tot}-\frac{\tau^{2}}{2}\hat{H}_{tot}^{2}\right)\hat{V}_{i,n}\left(1+i\tau\hat{H}_{tot}-\frac{\tau^{2}}{2}\hat{H}_{tot}^{2}\right)\right\rangle_{\rho_{SE}}\nonumber\\
&=&\left\langle\hat{V}_{i,n}\right\rangle_{\rho_{SE}}-i\tau\left\langle\left[\hat{H}_{tot},\hat{V}_{i,n}\right]\right\rangle_{\rho_{SE}}
+\frac{\tau^{2}}{2}\left\langle\left[\hat{H}_{tot},\left[\hat{V}_{i,n},\hat{H}_{tot}\right]\right]\right\rangle_{\rho_{SE}}\nonumber\\
&=&\left\langle\hat{V}_{i,n}\right\rangle_{\rho_{SE}}
+\frac{\tau\sqrt{\tau}}{2}\left\langle\left[\hat{V}_{i,n'},
\left[\hat{V}_{i,n},\hat{H}_{S}+\hat{H}_{E_{i}}\right]\right]\right\rangle_{\rho_{SE}},
\end{eqnarray}
where we have used the relation 
$\left\langle\left[\hat{H}_{tot},\hat{V}_{i,n}\right]\right\rangle_{\rho_{SE}}=0$.
The work current can then be formulated as
\begin{eqnarray}
\dot{W}^{com}&=&\lim_{\tau\rightarrow 0}\left(\Delta W/\tau\right)\nonumber\\
&=&\lim_{\tau\rightarrow 0}\left(\frac{1}{\tau\sqrt{\tau}}\sum_{i=h,c}\sum_{n=1}^{N}\left[\left\langle \hat{V}_{i,n}\right\rangle_{\rho_{SE}}-\left\langle \hat{V}_{i,n}\right\rangle_{\rho'_{SE}}\right]\right)\nonumber\\
&=&\lim_{\tau\rightarrow 0}\left(\frac{1}{\tau\sqrt{\tau}}\sum_{i=h,c}\sum_{n,n'=1}^{N}
\left[-\frac{\tau\sqrt{\tau}}{2}\left\langle\left[\hat{V}_{i,n'},\left[\hat{V}_{i,n},\hat{H}_{S}+\hat{H}_{E_{i}}\right]\right]\right\rangle\right]\right)\nonumber\\
&=&-\frac{1}{2}\sum_{i=h,c}\sum_{n,n'=1}^{N}\left\langle\left[\hat{V}_{i,n'},\left[\hat{V}_{i,n},\hat{H}_{S}+\hat{H}_{E_{i}}\right]\right\rangle\right]\nonumber\\
&=&-\frac{1}{2}\sum_{i=h,c}\sum_{n=1}^{N}\left\langle\left[\hat{V}_{i,n},\left[\hat{V}_{i,n},\hat{H}_{S}+\hat{H}_{E_{i}}\right]\right\rangle\right]\nonumber\\
&&-\frac{1}{2}\sum_{i=h,c}\sum_{n\neq n'=1}^{N}
\left\langle\left[\hat{V}_{i,n'},\left[\hat{V}_{i,n},\hat{H}_{S}+\hat{H}_{E_{i}}\right]\right\rangle\right].
\end{eqnarray}
By identifying 
\begin{eqnarray}
\dot{W}^{com}_{loc}&=&-\frac{1}{2}\sum_{i=h,c}\sum_{n=1}^{N}\left\langle\left[\hat{V}_{i,n},\left[\hat{V}_{i,n},\hat{H}_{S}+\hat{H}_{E_{i}}\right]\right\rangle\right],\nonumber\\
\dot{W}^{com}_{non-loc}&=&-\frac{1}{2}\sum_{i=h,c}\sum_{n\neq n'=1}^{N}
\left\langle\left[\hat{V}_{i,n'},\left[\hat{V}_{i,n},\hat{H}_{S}+\hat{H}_{E_{i}}\right]\right\rangle\right],
\end{eqnarray}
we finally obtain $\dot{W}^{com}=\dot{W}^{com}_{loc}+\dot{W}^{com}_{non-loc}$.

In the collision model, the heat related to bath $E_i$ is defined as
\begin{equation}
\Delta Q_{i}=\left\langle\hat{H}_{E_{i}}\right\rangle_{\rho_{SE}}-\left\langle\hat{H}_{E_{i}}\right\rangle_{\rho'_{SE}}.
\end{equation}
The expectation value of $\hat{H}_{E_{i}}$ with respect the evolved state $\rho'_{SE}$ can be obtained as 
\begin{eqnarray}
\left\langle \hat{H}_{E_{i}}\right\rangle_{\rho'_{SE}}&=&\left\langle\hat{U}\left(\tau\right)\hat{H}_{E_{i}}\hat{U}^{\dagger}\left(\tau\right)\right\rangle_{\rho_{SE}}\nonumber\\
&=&\left\langle e^{-i\tau\hat{H}_{tot}}\hat{H}_{E_{i}}e^{i\tau\hat{H}_{tot}}\right\rangle_{\rho_{SE}}\nonumber\\
&=&\left\langle\left(1-i\tau\hat{H}_{tot}-\frac{\tau^{2}}{2}\hat{H}_{tot}^{2}\right)\hat{H}_{E_{i}}\left(1+i\tau\hat{H}_{tot}-\frac{\tau^{2}}{2}\hat{H}_{tot}^{2}\right)\right\rangle_{\rho_{SE}}\nonumber\\
&=&\left\langle\hat{H}_{E_{i}}\right\rangle_{\rho_{SE}}
-i\tau\left\langle\left[\hat{H}_{tot},\hat{H}_{E_{i}}\right]\right\rangle_{\rho_{SE}}
+\frac{\tau^{2}}{2}\left\langle\left[\hat{H}_{tot},\left[\hat{H}_{E_{i}},\hat{H}_{tot}\right]\right]\right\rangle_{\rho_{SE}}\nonumber\\
&=&\left\langle\hat{H}_{E_{i}}\right\rangle_{\rho_{SE}}
-\frac{\tau}{2}\left\langle\left[\hat{V}_{i,n'},\left[\hat{V}_{i,n},\hat{H}_{E_{i}}\right]\right]\right\rangle_{\rho_{SE}}.
\end{eqnarray}
${}$\newline
The heat current is thus determined as 
\begin{eqnarray}
\dot{Q_{i}}&=&\lim_{\tau\rightarrow 0}\left(\Delta Q_{i}/\tau\right)\nonumber\\
&=&\lim_{\tau\rightarrow 0}\left(\sum_{n=1}^{N}\left[\left\langle \hat{H}_{E_{i}}\right\rangle_{\rho_{SE}}-\left\langle {H}_{E_{i}}\right\rangle_{\rho'_{SE}}\right]/\tau\right)\nonumber\\
&=&\lim_{\tau\rightarrow 0}\left(\frac{1}{\tau}\sum_{n,n'=1}^{N}
\frac{\tau}{2}\left\langle\left[\hat{V}_{i,n'},\left[\hat{V}_{i,n},\hat{H}_{E_{i}}\right]\right]
\right\rangle_{\rho_{SE}}\right)\nonumber\\
&=&\frac{1}{2}\sum_{n,n'=1}^{N}\left\langle\left[\hat{V}_{i,n'},\left[\hat{V}_{i,n},
\hat{H}_{E_{i}}\right]\right]\right\rangle_{\rho_{SE}}\nonumber\\
&=&\frac{1}{2}\sum_{n=1}^{N}\left\langle\left[\hat{V}_{i,n},
\left[\hat{V}_{i,n},\hat{H}_{E_{i}}\right]\right]\right\rangle_{\rho_{SE}}\nonumber\\
&&+\frac{1}{2}\sum_{n\neq n'=1}^{N}
\left\langle\left[\hat{V}_{i,n'},\left[\hat{V}_{i,n},\hat{H}_{S}+\hat{H}_{E_{i}}\right]\right]
\right\rangle_{\rho_{SE}}.
\end{eqnarray}
By denoting 
\begin{eqnarray}
\dot{Q}_{loc}^{i}&=&\frac{1}{2}\sum_{n=1}^{N}\left\langle\left[\hat{V}_{i,n},
\left[\hat{V}_{i,n},\hat{H}_{E_{i}}\right]\right]\right\rangle_{\rho_{SE}},\nonumber\\
\dot{Q}_{non-loc}^{i}&=&\frac{1}{2}\sum_{n\neq n'=1}^{N}
\left\langle\left[\hat{V}_{i,n'},\left[\hat{V}_{i,n},\hat{H}_{S}+\hat{H}_{E_{i}}\right]\right]
\right\rangle_{\rho_{SE}},
\end{eqnarray}
we finally obtain $\dot{Q_{i}}=\dot{Q}_{loc}^{i}+\dot{Q}_{non-loc}^{i}$.

%

\end{document}